\documentclass[useAMS,usenatbib]{mn2e}
\usepackage{makeidx}
\usepackage{rotating}
\usepackage{graphicx}
\usepackage{graphics}
\usepackage{lscape}
\usepackage{aas_macros}
\usepackage{natbib}
\usepackage{multirow}

\newcommand{\gtsim}{\mbox{{\raisebox{-0.4ex}{~$\stackrel{>}{{\scriptstyle\sim}}$~}}}}

\newcommand{\kms}[0]{km~s$^{-1}$}
\newcommand{\mum}[0]{$\mu$m}

\newcommand{\Ha}[0]{\textrm{H}{\large $\alpha$}}
\newcommand{\Hb}[0]{\textrm{H}{$\beta$}}
\newcommand{\Pa}[0]{\textrm{Pa}{\large $\alpha$}}
\newcommand{\Pb}[0]{\textrm{Pa}{$\beta$}}
\newcommand{\wco}[0]{\textrm{W[CO]}}
\newcommand{\wpa}[0]{\textrm{W[Pa}$\alpha$\textrm{]}}
\newcommand{\wha}[0]{\textrm{W[H}$\alpha$\textrm{]}}
\newcommand{\msun}[0]{$\textrm{M}_{\odot}$}
\newcommand{\solar}[0]{\mbox{$_{\odot}$}}
\newcommand{\Av}[0]{$\textrm{A}_{\textrm{\scriptsize V}}$}
\newcommand{\Ak}[0]{$\textrm{A}_{\textrm{\scriptsize K}}$}
\newcommand{\Ebv}[0]{$\textrm{E}(B-V)$}
\newcommand{\Vs}[0]{$\textrm{V}_\star$}
\newcommand{\Vg}[0]{$\textrm{V}_\textrm{\scriptsize gas}$}
\newcommand{\Ss}[0]{$\sigma_\star$}

\newcommand{\Deg}[0]{$^{\circ}$}

\def\phn{\phantom{0}}

\title[The Central Region of M83]{The Central Region of M83
\footnote{Based on observations made with ESO Telescopes at the La Silla or Paranal Observatories under programme IDs: 64.N-0100(B), 265.B-5723(A) and 65.O-0612(A). Also based on observations made with the NASA/ESA Hubble Space Telescope, obtained from the data archive at the Space Telescope Institute. STScI is operated by the association of Universities for Research in Astronomy, Inc. under the NASA contract  NAS 5-26555.}}
\author[R. C. W. Houghton, N. Thatte]{R. C. W. Houghton$^{1}$\thanks{E-mail:
rcwh@astro.ox.ac.uk} and N. Thatte$^{1}$ \\
$^{1}$University of Oxford, Denys Wilkinson Building, Keble Road, Oxford, OX1
3RH }

\begin{document}

\date{}

\pagerange{\pageref{firstpage}--\pageref{lastpage}} \pubyear{2006}

\maketitle

\label{firstpage}

\begin{abstract}

We combine VLT/ISAAC NIR spectroscopy with archival HST/WFPC2 and
HST/NICMOS imaging to study the central 20\arcsec$\times$20\arcsec\ of
M83. Our NIR indices for clusters in the circumnuclear star-burst
region are inconsistent with simple instantaneous burst
models. However, models of a single burst dispersed over a duration of
6~Myrs fit the data well and provide the clearest evidence yet of an age
gradient along the star forming arc, with the youngest clusters
nearest the north-east dust lane. The long slit kinematics show no
evidence to support previous claims of a second hidden mass
concentration, although we do observe changes in molecular gas
velocity consistent with the presence of a shock at the edge of the
dust lane.

\end{abstract}



\section{Introduction}
M83 (NGC 5236) is a nearby (4.5 Mpc; \citealt{Thim03}) grand design spiral galaxy SAB(s)c \citep{deVauc91} in a group containing several bright galaxies including Centaurus A (NGC 5128) and NGC 4945. Within the central 20\arcsec\ lies the photometric peak (hereafter referred to as the nucleus) offset from the centre of symmetry of the outer bulge isophotes and the global gas kinematics \citep{Wolst88,Gallais91,Thatte00,Sakamoto04} and surrounded by a semicircular starburst region, some 3\arcsec\ - 8\arcsec\ from the nucleus. M83 has a close dynamical companion, NGC 5253 \citep{Rog74} which also contains recent star formation, albeit much more compact than that of M83. The closest passage of NGC 5253 occurred 1-2 Gyrs ago \citep{Rog74} and so the activity in these galaxies is likely driven by internal inflow of gas via the bar rather than from direct interaction as the clusters are mostly 1-10 Myrs in age \citep[hereafter H01]{Harris01}.

\subsection{Central star formation}\label{sec:csf}

\cite{Gallais91} first presented NIR images of sub-arcsecond resolution for the central region of M83 which lead to an improved understanding of the morphology: the region of young stars is composed of a bright circum-nuclear arc of star clusters, located $\sim$7\arcsec\ from the old nucleus. However, \citeauthor{Gallais91} also attempted to date the young clusters and (contrary to more recent work, discussed later) suggested that the clusters were younger on the \emph{south-east} end of the arc and older on the \emph{north-west} end (but with one anomalous young cluster at this old end). They were not alone in this conclusion: \cite{Heap93}, using WFPC photometry and IUE UV spectroscopy also suggested that the age gradient went from south-east to north-west, the youngest clusters being in the south-east end of the arc. \citeauthor{Heap93} also identified that the old stellar nucleus was \emph{not} a site of \Ha\ emission. \citet{Pux97} performed low resolution NIR spectroscopy along a single slit bisecting the starburst arc and found evidence for a radial age gradient. 
\citet{Elmegreen98} found a double ring structure in their (J-K) colour images of the centre of M83 and associated each ring with an inner Lindblad resonances (ILR).

However, when H01 made a comprehensive study of the star clusters using HST/WFPC2 narrow- and broadband data coupled with the latest Starburst99 population synthesis models \citep[hereafter SB99]{SB99}, they identified that the tangential age gradient along the starburst arc was in fact from \emph{north-west} to \emph{south-east} with the youngest clusters in the north-west (in direct conflict with earlier work but resolving the anomalous cluster in \citealt{Gallais91}). H01 also found the majority of the clusters in the arc to have ages in the range 5-7 Myrs and found evidence of age gradients perpendicular to the starburst arc with younger clusters along the periphery; they concluded such gradients were indicative of an inside-out propagation of star formation. Subsequently, \cite{Bresolin&Kennicutt03} found Wolf-Rayet features in UV spectra of clusters near the north-west end of the arc and dated this star formation at 5 Myrs (and the other regions at 7 Myrs), in rough agreement with the estimates in H01 for the similar regions.

\subsection{Nuclear Rings}\label{sec:intro:nuclear_rings}

Considering the evidence thus far presented, one would most likely conclude that the circumnuclear arc seen at the centre of M83 is a \emph{nuclear ring}. Nuclear rings are often found at the centres of barred spiral galaxies \citep{Buta86} as are nuclear spirals \citep{MP99,PM02} and their formation is thought to be very similar.
\cite{Athanassoula92} first showed how the orbits in barred galactic potentials lead to gas shocking on the leading edges to form dust lanes with the shocked gas then flowing along the potential to the centre. These shocks have been observed across dust lanes, both directly as velocity jumps in gas kinematics and indirectly from increased radio emission \citep[both discussed in detail by][]{Athanassoula92}. %

However, \citeauthor{Athanassoula92}'s early simulations didn't have the resolution to properly probe the central regions to determine exactly what happened to the inflowing gas; this was done by \citet{Piner95} using the similar models to \citeauthor{Athanassoula92} but with cylindrical coordinates for better spatial resolution at the centre. 
\citeauthor{Piner95} found that if the barred potential permitted two ILRs, the gas flow builds up between them to form a dense nuclear ring. This was also confirmed by \citet{Combes96} and \citet{Buta&Combes96} who went further to say that even in a potential that admits one ILR, the inflowing gas will build up there to form a ring. Given the rings and ILRs found by \citeauthor{Elmegreen98}, it appeared likely that the arc of young stars in M83 was formed from gas building up in this way and leading to star formation.

This ILR picture of nuclear ring formation has persisted despite recent work suggesting subtle differences. \citet{Regan&Teuben03} performed an exhaustive number of simulations like those of \citeauthor{Piner95} for various barred potentials and they conclude that it is not the presence of ILRs that dictate the formation of a nuclear ring, but the existence and properties of an \emph{orbit family}, namely the $x_2$ family: if this orbit family exists, so can a nuclear ring and the authors find \emph{an excellent correlation between the $x_2$ orbit of largest extent along the bar major axis and the radius of the nuclear ring}. If the $x_2$ family does not exist, the authors \emph{never} find a nuclear ring. There is some discussion as to definitions: an ILR is normally defined as the locus at which the pattern frequency of the bar equals the circular orbital frequency less half the epicyclic orbital frequency, e.g. $\Omega_b = \Omega - \kappa/2$, and in weak bars this same definition traces the largest extent of the $x_2$ orbit family. However, as \citeauthor{Regan&Teuben03} are careful to point out, in \emph{strong} bars with a large quadrupole moment and \emph{low} central mass concentrations, this is no longer the case and there can be regions where $\Omega_b < \Omega - \kappa/2$ but no $x_2$ orbits exist. 
\citet{Witold_paper2} has also performed high resolution hydrodynamical simulations of the central regions of barred galaxies and shows how nuclear rings and nuclear spirals are formed by the same hydrodynamical processes. 
Although \citeauthor{Witold_paper2} continues the old trend in matching ILRs with the nuclear spiral and ring phenomenon, he is careful to define the ILR as the largest extent of the $x_2$ orbit family, implying that the $x_2$ family exists when referring to an ILR. 
We will return to the results in \citet{Regan&Teuben03} and \citet{Witold_paper2}, particularly to their  predictions of the phase space distribution of the gas (velocity and density) in the ring and to the trajectories of stars on $x_2$ orbits, formed in the ring.

\subsection{Nuclear Disks}\label{sec:intro:nucleardisks}

There is clearly overwhelming evidence that gas can accumulate in the central regions of barred potentials, although the present high resolution hydrodynamical simulations do not predict if and how the gas may give rise to star formation or how the outflows from the young stars and supernovae would affect the gas flow on the ring. Lower resolution N-body simulations including stars, gas and \emph{star formation} and then combined hydrodynamical and N-body simulations with star formation were performed by \citet{Wozniak03} and \citet{Wozniak&Champavert06} after \citet{Emsellem01} observed drops in the velocity dispersion at the centres of double-barred spirals (also seen in barred spirals by \citealt{Chung&Bureau04}) and couldn't reproduce them in simple dynamical models. The simulations with gas flows, stellar motions and star formation showed that when gas shocks in a barred potential and flows to the centre, it can produce a central disk. This disk of cold gas then produces young stars which dominate the luminosity of the central region and give rise to a very low velocity dispersion because of the low random motion of the stars in the disk. An important difference between these simulations and the pure hydrodynamical ones discussed earlier is that the gas builds up in a disk and not a ring or spiral; subsequent star formation is similarly distributed in a disk \emph{and not a ring}. Whether this difference is a resolution problem or something more subtle remains to be seen, although if the surface density of the gas was higher in the central region of the disk, one would expect star formation to initiate there and propagate outwards, potentially leaving an outer ring of young stars as in \citet{KenneyCarlstromYoung93} and a \emph{radial} age gradient in the starformation (consistent with the observations of M83 made by \citeauthor{Pux97}, although H01 report younger star clusters around the periphery of the arc, both on the inner and outer edges).

\subsection{Polar rings and disks}
However, there are certain complications with the circumnuclear ring interpretation for the centre of M83. Most previous studies discuss an arc of young stars, not a complete ring, apart from \citet{Gallais91} who claim to observe a complete ring or ellipse of star clusters from their sub-arcsecond H-band image, and \citet{Elmegreen98} who found circum-nuclear rings in their (J-K) NIR images. The fact that complete ring structures are seen in the NIR and not in the optical is highly suggestive of an obscuration of the complete ring by dust; an idea supported by extinction estimates we present later in \S\ref{sec:results}. And there may be a simple explanation for the obscuration of one half of a ring and not the other: \citet{Sofue&Wakamatsu93} argue that the north-east dust lane is observed to warp out of the plane of the main disk as it approaches the nucleus and they go so far as to call it polar; a similar warp of the south-west dust lane on the other side of the galaxy might explain why it is also not visible. The hydrodynamical simulations of \citet{Athanassoula92,Piner95,Regan&Teuben03,Witold_paper2} are all in 2D and are unable to permit such warping out of the disk plane. We therefore have no theoretical predictions of gas inflow producing warped disks, spirals and rings, although we can see similar behaviour in other galaxies such as NGC~5383 \citep[as noted in][]{CarnegieAtlas}. If similar hydrodynamical processes occur in 3D as they do in 2D but the gas inflow can build up in a polar or inclined ring out of the plain of the galactic disk, we would certainly expect the star formation to do likewise and given a highly obscuring and dusty galactic disk, for the inclination of M83 to our line-of-sight we would only see the near side of the inner warped disk and an incomplete ring of young stars, as explained by \citeauthor{Sofue&Wakamatsu93}. However, \citeauthor{Sofue&Wakamatsu93} never comment on the plausibility of the galaxy's inclination giving rise to the obscuration of the dark south-west dust lane by the luminous bulge. 

\subsection{Interloping hidden mass concentrations}\label{sec:intro:interlopers}

After studying the centre of M83 with near infra-red (NIR) long-slit kinematics, \citet[][hereafter T00]{Thatte00} initially suggested the possibility of a second nucleus or mass concentration after the stellar kinematics across the principal nucleus (Slit E in Fig. \ref{fig:slits_knots}, cutting the arc of young stars and the `bridge' identified in \citealt{Gallais91}) revealed two peaks in the velocity dispersion --- one associated with the nucleus and another which was 2\farcs7 SW of the nucleus. Assuming that the stars are dynamically relaxed in the gravitational potential, the secondary peak in the dispersion is best explained by invoking the presence of a second dynamically hot but obscured mass concentration (T00). Subsequently, \citet{Mast06} performed optical integral field spectroscopy (IFS) on the central region and linked a strong velocity gradient in the \Ha\ velocity map (3\farcs9$\pm$0\farcs5 W of the nucleus) to the location of the proposed second obscured mass concentration. However, the same authors \citep[hereafter D06]{Diaz06a} then performed IFS on a coincident but larger region, with better spatial resolution and in the near infrared (NIR). They then found that the largest gradient in the \Pb\ velocity field was in fact some 7\arcsec\ WNW of the nucleus and so adopted this position as a likely location for a mass concentration \citep[][hereafter D06b]{Diaz06b}. The most recent work \citep{Diaz07} now reports three mass concentrations: the visible nucleus and a further \emph{two hidden} mass concentrations at two different locations, both apparently more massive than the visible nucleus. Despite the inconsistency between \Ha\ and \Pb\ kinematics (c.f. Fig. 5 of \citeauthor{Mast06} and Fig. 3 of D06a), gas kinematics in general are well known to exhibit non-gravitational effects \citep{KormRich95} and the strong velocity gradients observed in the gas kinematics could be caused by hydrodynamical effects (e.g. outflows, shocks and spiral density waves) rather than mass concentrations. 

Nevertheless, the scenario of D06(a,b) is appealing, given the theory of dynamical friction in a gaseous medium as presented by \citet{Ostriker99} and the numerical implementation. Simulations of merging black holes (BHs) in a gaseous medium \citep[e.g.][]{Escala04,Kim&Kim07} reproduce density wakes tantalisingly similar to the density distributions of the aforementioned nuclear spirals and rings; one would be hard pressed to distinguish between Fig. 5 of \citet{Escala04} and Fig. 6 of \citet{Witold_paper2}. Furthermore, Fig. 2 in \citet{Escala04} bears a remarkable resemblance to the scenario seen at the centre of M83, assuming that the high density gaseous wake can initiate star formation. As pointed out by D06(a,b), invoking this scenario nicely explains the age gradient of H01 (and the average age gradient of the H01 data presented in D06b) as well as the off-centred old nucleus.

However, some important queries remain unanswered. If the gradient in the gas kinematics is caused by the gravitational potential, one should see the same effect in the stellar kinematics which are mostly unaffected by shocks or outflows and accurately trace the gravitational potential (although if the shock and accumulation of gas is significant, it may begin to dominate over the smooth potential of the disk and bulge). H01 also date some of the youngest clusters in the arc to be \emph{beyond} the present location of the interloping mass of D06(a,b). If they are related to the arc of young stars, it is not clear how they got to be there: the high density wake created by the interloping masses in \citet{Escala04} and \citet{Kim&Kim07}, which we assume leads to star formation and an age gradient, is always trailing and never preceding the interloping mass. Furthermore, the clusters nearest the supposed location of the interloper have been dated at 5 Myrs (H01 and \citealt{Bresolin&Kennicutt03}); this is larger than one would expect given the age gradient, the dynamical timescale and their small distance from the interloper (disscussed again in \S\ref{sec:discussion}). 

Perhaps more importantly, there is very little evidence of any merger event having occurred recently (the last 50 Myrs) in M83. The most prominent tidal tail of M83 was first published by \citet{Malin97}: it is roughly 27\arcmin\ in length according to this exposure, subtends around 60\Deg\ in the north by north-west and is estimated to be between 15\arcmin~(20kpc) and 20\arcmin~(26kpc) from the centre of M83. Clearly, such an arc is well outside the main galatic disk, defined either as the extent of the HII emission or the Holmberg radius \citep[5\farcm1 and 7\farcm3 respectively, in][]{Thilker05}. However, this tidal tail is within the HI distribution as seen in \citet{Park01} and \citet{Miller&Bregman04}; in fact, both these HI studies see a north-western arc of HI emission. The figures in \citet{Park01} clearly show this HI arc to extend right over from the north-west to the north-east and they estimate a radius of 25\arcmin\ from the centre of M83. This is approximately the same location as the stellar arc found by \citet{Malin97} and rescaling and overlaying the two images reveals that they are indeed co-spatial near the north-west tip of the stellar arc but diverge to the north-east with the HI at a larger radius than the stellar feature. Such features are so far outside the main stellar disk of M83 that the disrupted source is unlikely to have fallen via dynamical friction to the bottom of the potential, but is more likely to 
contribute to the outer halo. In fact, on visualising the aforementioned overlay, one is tempted to assign the disturbed outer HI structure seen in \citet{Park01} to the capture of a satellite which we now see as the \citeauthor{Malin97} stellar arc.
Moving closer to the galactic disk, \citet{Thilker05} report GAIA observations of two bisymmetric filaments or arms just outside the main HII disk (5\farcm1), one extending roughly north from the east of the disk and the other extending roughly south from the west of the disk, both approximately 15\arcmin\ in length. Such beautyfully symmetric filaments of star formation are unlikely to be induced by a merger event, but rather by the remaining HI envelope as the authors propose. The main stellar disk of M83 is remarkably regular (as seen in Fig. \ref{fig:eso_danish}) with only one potential blemish  (see \S\ref{sec:ack}): radio studies by \citet{Cowan94} and \citet{Maddox06} identify a linear sequence of 3 sources to the north-east of the central region. However, in \citet{Maddox06}, the central source of the three is identified to be coincident with one of the x-ray sources of \citet{Soria&Wu03} and the authors conclude that this linear structure is a background radio galaxy with two lobes of emission from jets and a central galaxy source.\footnote{Is is worth noting that there are two other sources roughly along the same linear structure in the south-east, sources 32 and 36, which both have x-ray counterparts leading to them being identified as x-ray binaries (XBRs) in \citeauthor{Maddox06}'s study.}. 

Finally, the CO map of \citet{Sakamoto04} is regular and symmetric both in its density and its velocity, which is unlikely to be the case if a recent merger had disturbed the gas sufficiently to ignite star formation in its wake.

\subsection{This Study}\label{sec:intro:thisstudy}

In this paper we report on the results of combining HST photometry and ESO VLT NIR long-slit spectroscopy at the centre of M83. We address the claims of T00 and D06(a,b) regarding the presence of a second obscured mass concentration by looking for further dynamical signatures in the stellar kinematics at the same locations. In particular, to help identify a cause for the gradient in the gas kinematics seen in D06a, we observe the NIR stellar kinematics at the same location. We also build on the study of H01 and compare optical and NIR indices of clusters in the circumnuclear arc (some of which were undetected in H01) with models for the stellar populations. Although H01 showed that an age gradient existed along the arc from north-west to south-east with the youngest clusters in the north-west, the reddening vector in this study parallels the tracks in the two-colour diagram for ages of 5-10 Myrs, complicating the age analysis \citep{Ryder05}, particularly because the extinction estimates were derived from the \Hb:\Ha\ decrement which suffers from low \Hb\ flux, poor penetration and a small wavelength range. NIR indices can probe deeper into potentially obscured clusters, giving a more representative measure of the ages. Furthermore, clusters can be heavily extincted and therefore overlooked in the visible, which is particularly relevant for very young clusters ($<$ 5 Myrs) which may be dustier and only have a weak intrinsic stellar continuum \citep{Ryder05}; this is less of a problem in the NIR, but we also present and correct our data using an extinction map derived from the \Ha:\Pa\ decrement, benefitting from more flux, a larger wavelength range and the penetrating \Pa\ data. 

The structure of this paper is as follows: \S\ref{sec:obs_and_reduction} describes the reduction and homogenisation of data from different instruments; \S\ref{sec:data_analysis} describes the analysis of the data; we briefly discuss some important details regarding the error analysis in \S\ref{sec:errors}; \S\ref{sec:results} presents the results which are subsequently discussed in \S\ref{sec:discussion} while \S\ref{sec:conc} concludes.


\section{Observations and Data Reduction}
\label{sec:obs_and_reduction}

We present unpublished VLT/ISAAC data and archival HST/NICMOS and HST/WFPC2 data. The observations and data reduction for each instrument are discussed separately below, followed by details of how they were homogenised. Fig. \ref{fig:slits_knots} illustrates the slit positions relative to the HST data.

\begin{figure*}
  \includegraphics[width=\textwidth]{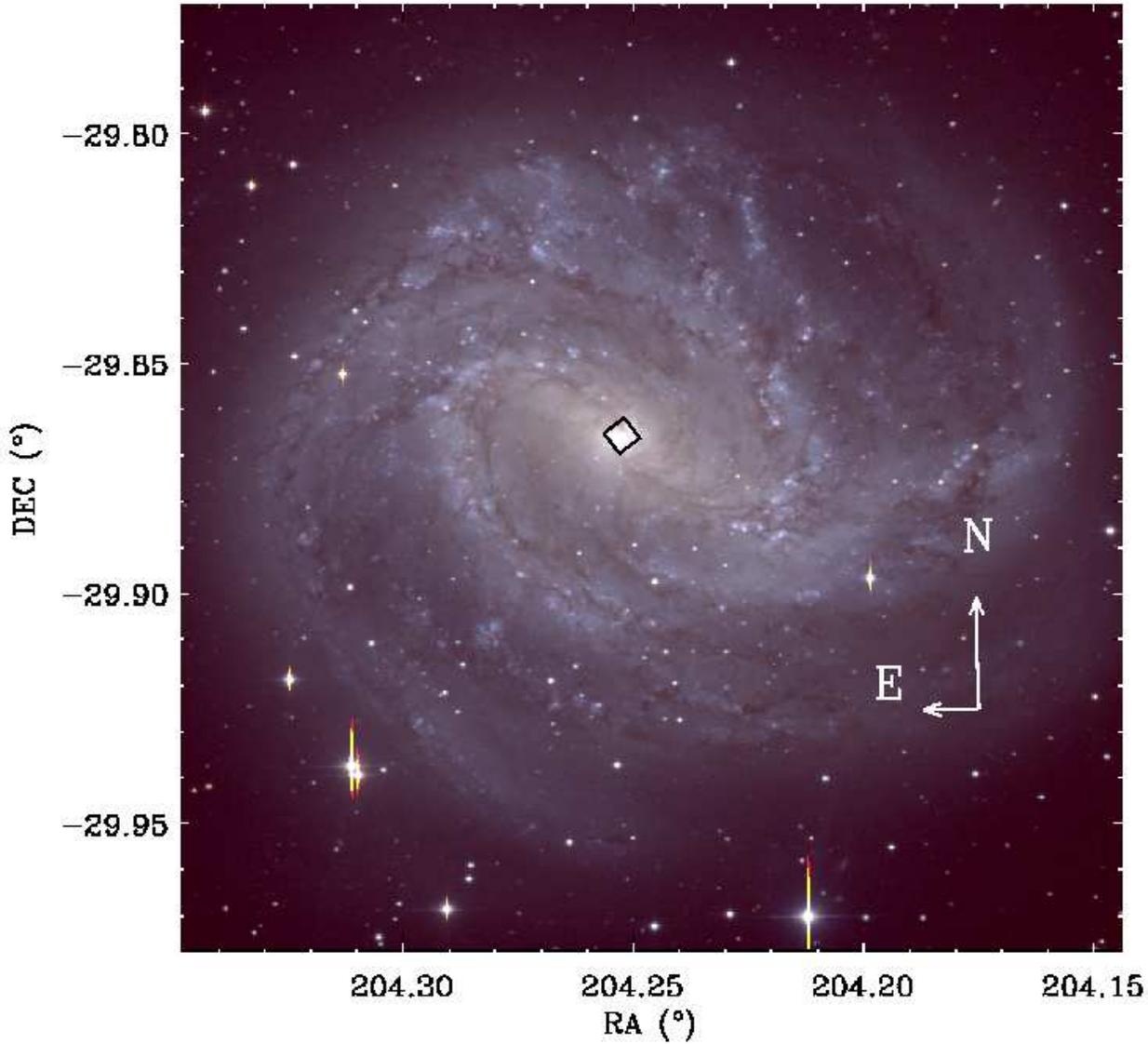}
  \caption{A BVR false-colour image of the spiral
  galaxy M83 (using logarithmic intensity scaling), with the footprint of
  the homogenised HST/WFPC2 and HST/NICMOS images overlaid in black. The images were observed using the ESO Danish 1.54m telescope \citep{Larsen99}}
    \label{fig:eso_danish}
\end{figure*}

\subsection{VLT/ISAAC Spectroscopy}
\label{sec:isaacred}

New K band long-slit spectroscopy along 4 different positions (A, B, C and D in Fig. \ref{fig:slits_knots}) was observed in late 2000 and we combine this with the original major axis data of T00 (E in Fig. \ref{fig:slits_knots}) to give a total of 5 slit positions in the central region of M83 observed with the VLT/ISAAC spectrograph using the 0\farcs6 slit (0\farcs147 pix$^{-1}$ plate scale) and an SK filter to give a wavelength range of 2.239\micron\ - 2.362\micron\ at a resolution $(\lambda/\Delta\lambda)$ of 4400. Observations were made using the ABBA technique, removing the need for separate sky exposures. Table \ref{tab:isaacexp} summarises other details regarding the exposures.

\begin{table}
  \begin{centering}
    \begin{tabular}[t]{cccccc}
        \hline
         Date    & ID &PA (\Deg)    &  Time   & DIMM  & Data \\
	 D/M/Y   &    &             &  (mins) & FWHM  & FWHM   \\    
	\hline
	 11/07/00     & [A] &126.95   &  40 (5) &    1\farcs02 &    0\farcs6  \\
         13/07/00     & [A] &126.95   &  40 (5) &    0\farcs52 &    0\farcs4  \\
	 17/08/00     & [B] &124.89   &  30 (5) &    0\farcs74 &    0\farcs5  \\
	 18/08/00     & [C] &118.13   &  30 (5) &    0\farcs61 &    0\farcs5  \\
	 19/08/00     & [D] &171.26   &  30 (5) &    2\farcs22 &    1\farcs2  \\
	 21/03/00     & [E] & 51.15   &  15 (5) &    0\farcs71 &    0\farcs4  \\
	\hline
    \end{tabular}
    \caption{A summary of the VLT/ISAAC K-band long-slit
    spectroscopy. The time column gives the total exposure time of all
    frames and, in parenthesis, the exposure of the individual
    frames. The last two columns compare the mean seeing reported by
    the DIMM and the seeing estimated directly from the data
    (\S\ref{mergeHST+ISAAC}). The ESO program IDs for the March,
    July and August exposures are 64.N-0100(B), 265.B-5723(A) and
    65.O-0612(A), respectively.}
    \label{tab:isaacexp}
  \end{centering}
\end{table}

A number of kinematic templates were observed with the same
instrumental setup on previous programmes (ESO 64.N-0100 and
65.N-0577). However, this template library lacked supergiants which
are typical of a starburst population. Consequently, we supplemented
the library with a number of supergiant observations also made with
ISAAC but using the 0\farcs3 slit (ESO 68.B-0530, 67.B-0504). To
homogenise the library, the latter (of higher spectroscopic
resolution) were convolved down to match the resolution of the
0\farcs6 slit templates. The 0\farcs3 slit templates also had a
shifted wavelength range (2.249\micron - 2.373\micron), so both the 0\farcs3 and 0\farcs6
slit templates were cropped to the common wavelength range. Our final library consisted of 6 giants (K1--M6) and 14 supergiants (K5--M5).

Reduction of the long slit data mainly followed the standard
technique, as detailed in the VLT/ISAAC
user-manual\footnote{http://www.eso.org/instruments/isaac/doc/} with
use of IRAF\footnote{http://iraf.noao.edu/} and
ECLIPSE\footnote{http://www.eso.org/projects/aot/eclipse/} routines.
Minor differences to the above are summarised below. 

The raw data frames were corrected for the odd-even effect prior to
any other processing using the ECLIPSE Jitter routine. Persistent bad
pixels were detected in the flat field images and interpolated over
during flat fielding. Subsequently, median filtering was used to
detect (but not replace) cosmic rays to make a mask for each data
frame. After rectification of the data to a uniform wavelength and
spatial grid, the frames were aligned and stacked by fitting a
Gaussian to a star cluster in the flux profile of each
frame. Crucially, the previous masks were also rectified and shifted
in the same manner to identify contamination into neighbouring
pixels. All contaminated pixels were then excluded when combining the
stacked data frames. Additionally, the data underwent an iterative
clipping (at $5\sigma$) as a precaution to eliminate unidentified
bad/hot pixels. It is possible to simply remove cosmic rays prior to
rectification and alignment by interpolation of neighbouring pixels,
but the method employed here interpolates from additional realisations
of the same pixel, which we prefer wherever possible.

Telluric standard stars (HD119013 and HD118187) were also observed
each night with the same spectrograph setup to correct the other
observations for atmospheric absorption, using standard correction
techniques \citep{omo93}.

We found that the absolute wavelength calibration of the observations
varied each night and was not corrected for by the ARC exposures. To
obtain an accurate calibration (the importance of which is described
in \S\ref{sec:wfpc2_images}), we used the OH emission line at
$\lambda_\textrm{vac}22460.264\textrm{\AA}$ in the unsubtracted frames
as a reference point and corrected accordingly. We further applied a
heliocentric velocity correction for each night to correct to an
absolute velocity scale.

Slit A was observed over two nights (Table \ref{tab:isaacexp}) but the
seeing varied significantly between them. For all subsequent analysis,
we chose to use the data with the best seeing (that of 13/07/00); the
increase in the signal-to-noise ratio (SNR) from adding an additional
night was marginal and significantly degraded the spatial resolution
(increasing cluster contamination). Problems with the arc lamp
observations (only the Argon lamp and not the Neon lamp was operating)
meant the wavelength calibration was also not as accurate for the
11/07/00 observations of slit A, further motivating our decision to
neglect them. The position of slit B, although very similar to that of
A, covers a slightly different region and is better centred on a few
knots, as can be seen in Fig. \ref{fig:slits_knots}. We therefore do
not combine the data with Slit A, but analyse it separately.

\subsection{HST/NICMOS Images}

\begin{table}
\begin{centering}
\begin{tabular}{cr@{~}lcccc}
\hline
  Date & \multicolumn{2}{|c|}{Filter} & Time  & Mean $\lambda$ & Bandpass  \\
 D/M/Y &    &    &  (s)  & (\AA)          & (\AA)  \\
\hline
\multicolumn{6}{|c|}{ WFPC2 (Proposal ID: 8234, PI: Calzetti)}\\
\hline
 25/04/2000 & F300W&(U)& 700  & 3014 & 858.3  \\
 02/05/2000 & F487N&(\Hb)& 1100 & 4866 & 33.9   \\  
 25/04/2000 & F547M&(V)& 310  & 5488 & 637.9  \\
 02/05/2000 & F656N&(\Ha)& 600  & 6564 & 28.3   \\
 25/04/2000 & F814W&(I)& 237  & 8023 & 1472.8 \\ 
\hline
\multicolumn{6}{|c|}{NICMOS (Proposal ID: 7218, PI: Rieke)}\\
\hline
 16/05/1998 & F187N&(\Pa)& 160 & 18740 & 194.8  \\ 
 16/05/1998 & F190N&(cont.)& 160 & 19003 & 184.0  \\
 16/05/1998 & F222M&(K)& 176 & 22181 & 1479.4 \\
\hline
\end{tabular}
\caption{A summary of the HST observations. Mean wavelengths and
bandpasses are calculated with \emph{bandpar} in the STSDAS synphot
package (AVGWV and RECTW, respectively). }
\label{tab:hstexp}
\end{centering}
\end{table}

The central $20\arcsec\times20\arcsec$ of M83 were observed with
HST/NICMOS in the narrow band filters F187N (\Pa), F190N (\Pa\
continuum) and F222M (approximately K-band), the details of which are
shown in Table \ref{tab:hstexp}. Fig. \ref{fig:eso_danish} illustrates the NICMOS
footprint on a false-colour composite \citep[][observed with the ESO Danish 1.54m telescope and available on NED]{Larsen99}. All exposures were
made with the
NICMOS 2 camera with an approximate plate scale of 0\farcs075 per
pixel (see \S\ref{sec:mergeNICMOS+WFPC2}). The data were reduced
using the STSCI calibration pipeline. However, for exposures observed
in non-chopping mode (i.e. without background exposures), the latter
stage of the pipeline (calnicb) assumes a sparsely populated field and
is configured to estimate and subtract off the background level from
the same exposure. It does this by identifying and masking sources
(groups of 2 or more pixels $4.5\sigma$ higher than the image median)
and using the remaining sigma clipped median as the background level.
It is not difficult to see that for our non-chopped observations of
the centre of M83, this process will lead to gross over subtraction of
the background. Consequently, the value estimated for the background
by the pipeline is added back onto the images. They are then flux
calibrated using the photometric zero-point PHOTFLAM calculated by the
STSCI pipeline and RECTW bandpasses (for narrow-band images) from the
STSDAS \emph{bandpar} routine. The background of each mosaiced
exposure is estimated from a suitable blank region near the edge of
the frame which is then subtracted.

\subsection{HST/WFPC2 Images}
\label{sec:wfpc2_images}

\begin{table}
\begin{centering}
\begin{tabular}{ccccc}
\hline
 $V_\textrm{rec}$  & \Ha\ Trans.  & \multirow{2}{*}{${\rm [NII]\over
H{\alpha}}$}  & \Hb\ Trans. & \multirow{2}{*}{${\rm F656N \over F487N}$}\\
  (\kms)    & Ratio (\%)   &                      & Ratio (\%)  & \\
\hline
   500     &      82.6   & 24.6      & 96.8   & 0.83    \\
   512     &      80.6   & 25.2      & 96.5   & 1.03    \\
   550     &      72.8   & 27.9      & 95.8   & 1.12    \\
   600     &      59.5   & 34.0      & 94.8   & 1.32    \\

\hline
\end{tabular}
\caption{The effect of [NII] contamination and filter transmission
with recession velocity $V_\textrm{rec}$ for a Gaussian \Ha\ emission
line of 150\kms\ FWHM. The columns represent: the recession velocity
of the gas $V_\textrm{rec}$; the \Ha\ transmission at $V_\textrm{rec}$
as a percentage of the transmission at $V_\textrm{rec}=0$\kms; the
[NII] contamination as a percentage of the \Ha\ emission; the \Hb\
transmission at $V_\textrm{rec}$ as a percentage of the transmission
at $V_\textrm{rec}=0$\kms; the ratio of the observed (uncorrected)
fluxes in the F656N (\Ha) filter and the F487N (\Hb) filters with
$V_\textrm{rec}$ (note the effect this may have on extinction
estimates using these filters).}
\label{tab:filt_trans}
\end{centering}
\end{table}

The central $\index{\footnote{}}36\arcsec\times36\arcsec$ of M83 was observed with WFPC2 in
the filters F814W, F547M, F300W, F487N (\Hb) and F656N (\Ha),
details of which are given in Table \ref{tab:hstexp}.

The data were reduced using the STSCI calibration pipeline. In
addition, the STSDAS routine \emph{wfixup} was used to interpolate
over known bad pixels. Flux calibration was performed using the
PHOTFLAM photometric zero point calculated by the STSCI pipeline and
RECTW bandpasses (for narrow-band images) from the STSDAS
\emph{bandpar} routine.

The transmission curves of the F656N and F814W filters were obtained
from the STSDAS \emph{calcband} routine. Using the optical spectra of
\citet{Storchi95}, the [NII] and \Ha\ emission lines were fitted (with
Gaussians) and subtracted; subsequently applying the F656N and F814W
filter curves gave the ratio of the continuum flux in each of the
filters. Background subtraction for the F656N image was then performed
using the F814W image multiplied by this normalisation factor (as done
by H01).

There are two systematic errors associated with the background
subtracted F656N observations which required correction. Firstly, the
F656N filter is broad enough to include [NII] emission in addition to
the \Ha\ emission. Secondly, because of the recession velocity of M83,
\Ha\ lies near the edge of the bandpass, reducing the observed
\Ha. These two systematic errors are also correlated: the recession
velocity of M83 shifts the $\lambda_\textrm{vac}6585.27\textrm{\AA}$
[NII] line out of the bandpass, significantly reducing the
contribution from this line. However, it also shifts the
$\lambda_\textrm{vac}6549.86\textrm{\AA}$ [NII] line into the
bandpass, \emph{increasing} the contribution.

Correcting for these two effects is not trivial. Using the optical
spectra of \citet{Storchi95}, one may attempt to correct for both, but
the spectra are low resolution (FWHM $\sim$ 10\AA\ $\sim$ 500\kms)
causing the lines to be significantly broadened. Such instrumental
broadening hinders our calculations as it is not possible to reliably
estimate the change in transmission for a recession velocity of
500\kms\ when the line itself is 500\kms\ in width.
Therefore we fit the \Ha\ and [NII] lines in the \citet{Storchi95}
spectra with Gaussians to estimate their relative fluxes and then
reproduce the spectrum with lines in the same flux ratios but each
with a FWHM=150\kms\ (in agreement with the ionised gas kinematics of
D06). Using this (continuum subtracted) spectrum, we calculate the
contribution of the [NII] lines and the transmission of the \Ha\ line
in the F656N image for a range of recession velocities. Table
\ref{tab:filt_trans} summarises our findings.

Although D06 quote absolute ionised gas velocities ranging between
550\kms\ and 650\kms, the stellar velocity field is not given, so we
have no knowledge of whether the ionised gas is actually at a
different velocity to the stars or if the absolute wavelength scale is
in conflict with the recent determination of $V_{rec} = 512$\kms by
\citet{Kori04}. This latter value is in good agreement with the
average recession velocity found in the HyperLEDA database (507\kms).
We took considerable pains to calibrate our ISAAC data to an absolute
heliocentric velocity scale (\S\ref{sec:isaacred}). The error
weighted mean stellar velocity along all slits was found to be
520\kms. However, the error weighted mean molecular Hydrogen velocity
was found to be 512\kms. We therefore adopt the value of
\citet{Kori04} for the ionised gas and correct accordingly. Quite by
chance, the [NII] contamination and bandpass edge effects almost
counteract each other at this velocity, leaving only an overall
scaling of 0.99 for the F656N image.

The spectra of \citet{Storchi95} are averages over the central region
and although we attempt to account for the non-uniform gas
velocity across the FoV with a systematic error (see \S\ref{sec:errors:wha}), we
might also expect variations in the [NII]:\Ha\ flux ratio. Thus our
correction in this matter is only to first order.

The effect of recession velocity is smaller for the WFPC2 F487N (\Hb)
filter (Table \ref{tab:filt_trans}) and is non-existent for the NICMOS
F187N (\Pa) filter so the F487N image received a minor flux correction
of 1.036 in accordance with Table \ref{tab:filt_trans} (neither filter
is contaminated by other emission lines).

\subsection{Merging HST/WFPC2 data with HST/NICMOS data}
\label{sec:mergeNICMOS+WFPC2}

HST narrow band images exist in \Ha\ (WFPC2) and \Pa\ (NICMOS). In
order to homogenise these two data sets, it was necessary to match the
pixel sampling, point spread function (PSF), position angle (PA) and
alignment. Fig. \ref{fig:eso_danish} illustrates the footprint of the
homogenised HST images on a false-colour composite.

A well known feature of NICMOS detector is the rectangular and time
variable plate scale. However, the feature is well documented and we
interpolated the plate scale at the time of our observations from the
plate scale records available on the NICMOS STSCI website. At the time
of the \Pa\ observations, the plate scale was 0\farcs0759788 $\pm$ 0\farcs0000025 and
0\farcs0752962 $\pm$ 0\farcs0000025 in \emph{x} and \emph{y}, respectively (we assume that
the plate scale of the WFPC2 detector is constant and adopt the value
of 0\farcs04554 quoted in \citealt{Holtzman95}). We re-sampled the
original 261$\times$61 NICMOS images to uniform 261$\times$61 arrays with a
0\farcs0752962 pix$^{-1}$ plate scale in \emph{x} and \emph{y}. Flux
was conserved throughout.

The WFPC2 images were rotated to the same PA of the NICMOS images
(using the ORIENTAT keyword) and convolved with a Gaussian PSF of
$\sigma=1.355$ pixels (the FWHM of a diffraction limited PSF for a
2.4m aperture at 1.876\micron\ is 1.503 pixels; if we assume that this
Airy disk can be approximated by a Gaussian, by adding $\sigma$s in
quadrature we reproduce the FWHM of the NICMOS F187N PSF). The rotated
and convolved 800x800 planetary camera images were then re-sampled to
the same plate scale as the re-sampled NICMOS images.

\begin{center}
  \begin{figure*}
  \includegraphics[width=\textwidth]{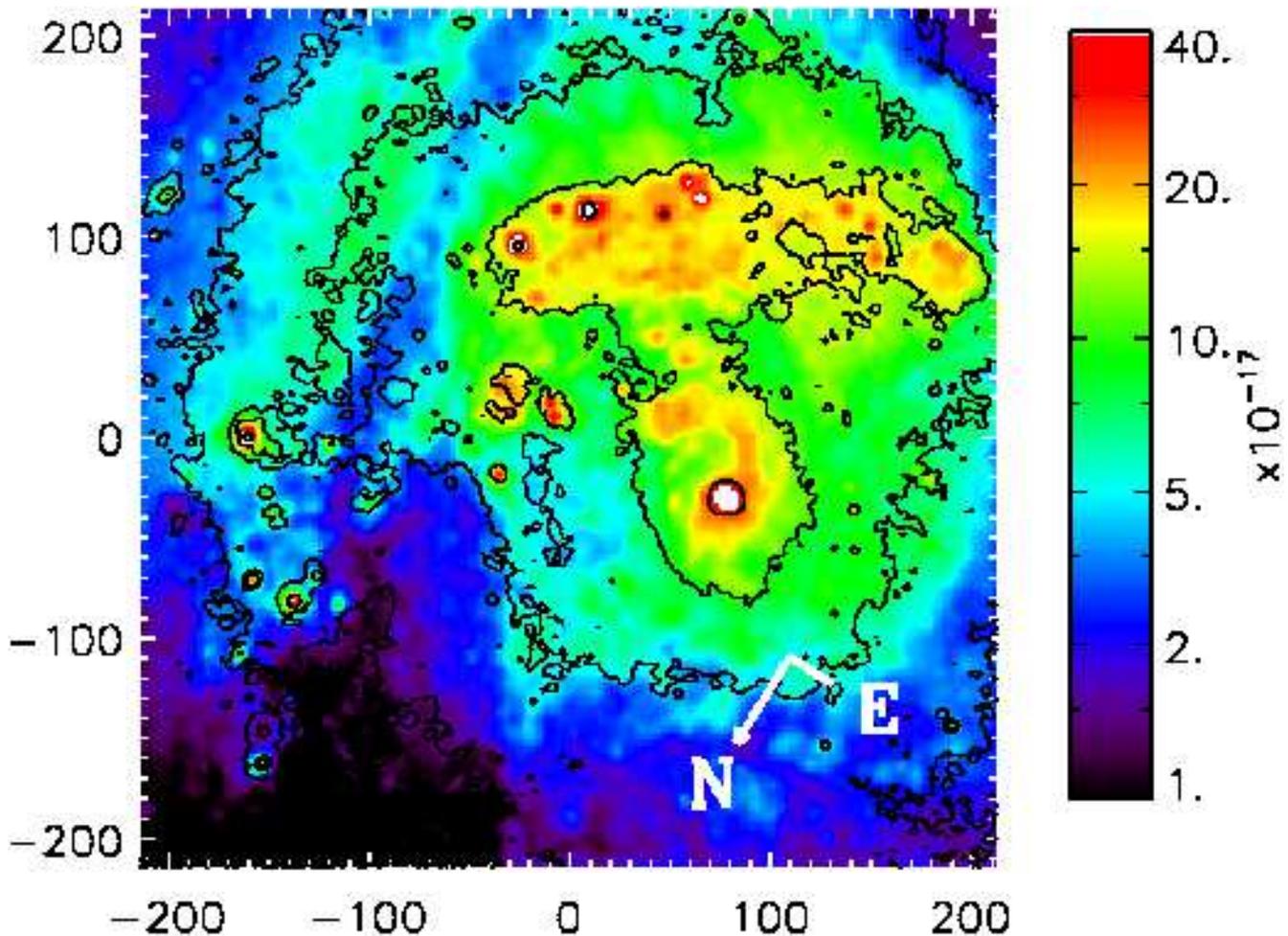}
  \caption{The homogenised WFPC2 F814W image, shown with logarithmic
  colour map scaling. Over-laid are contours of the homogenised NICMOS
  F222M image at levels $(1.3\times10^{-15}, 1.8\times10^{-15},
  2.5\times10^{-15}, 7.4\times10^{-15})$. The axes are scaled in
  pc and all fluxes are given in ergs s$^{-1}$ cm$^{-2}$. Note
  the accuracy of the spatial scaling and alignment between the WFPC2
  and NICMOS data. The footprint of this image is shown in Fig. \ref{fig:eso_danish}.}
    \label{fig:wn_overlay}
  \end{figure*}
\end{center}

Finally, we shifted and aligned the rescaled WFPC2 F814W image to the
NICMOS F222M image by maximising the cross-correlation function. We
applied the same shift to the other WFPC2 images and extracted 261$\times$61
images corresponding to the NICMOS FoV. The end result was a
remarkably accurate scaling and aligning of WFPC2 and NICMOS data,
demonstrated in Fig. \ref{fig:wn_overlay}. The resolution of the final images (e.g. Figs. \ref{fig:wn_overlay}, \ref{fig:slits_knots}, \ref{fig:ext_map} and \ref{fig:cluster_ages}) is set by the NICMOS instrument and we estimate the FWHM to be around two pixels (0\farcs15).
%

\subsection{Synthetic Slits: merging space and ground based data}
\label{mergeHST+ISAAC}
In order to merge the ISAAC data with the NICMOS and WFPC2 datasets,
we extracted apertures from the HST data corresponding to the ISAAC
slit positions. As the NICMOS and WFPC2 datasets were already
homogeneous (\S\ref{sec:mergeNICMOS+WFPC2}), it remained to
reproduce the seeing conditions and locate and extract synthetic
slits.

To facilitate this procedure, a routine was written to minimise the
$\chi^2$ difference between the ISAAC long-slit flux profile (short-ward
of the CO bandhead at 2.3\mum) and a flux profile produced by
overlaying a 0\farcs6 slit aperture on the NICMOS F222M image,
incorporating sub-pixel shifting, rotation and Gaussian convolution to
emulate the seeing conditions. The problem was highly non-linear, with
many local minima which could stall the process. Thus the optimisation
required a good initial guess for the parameters (made by eye).

The principal result of this routine was the extraction of series of
synthetic slits from the HST images, aligned to the ISAAC data and
matched in spatial resolution. A by-product was an accurate estimate
for the seeing of the ISAAC ground-based data (Table
\ref{tab:isaacexp}) which allows for greater accuracy and less
contamination when extracting spectra of individual star clusters
(\S\ref{sec:isaac_analysis}). The reported seeing from the
Differential Image Motion Monitor (DIMM) did not always represent the
true seeing of our data, which is not surprising given that the DIMM
looks in a roughly fixed direction \citep[less than $30\deg$ from the
zenith]{ESODIMM} and at visible wavelengths; the ISAAC data was taken
at a different position on the sky, with a different airmass and in
the NIR (recall that seeing $\propto \lambda^{-0.2}$).


\section{Data Analysis}
\label{sec:data_analysis}

We wish to build on the study of H01 and compare optical and NIR
indices for clusters in the circumnuclear arc with stellar population
models and also address the claims of T00 and D06
regarding the presence of a second obscured mass concentration. The
first of these goals requires chemical and kinematic analysis of
individual clusters. The latter requires kinematic analysis along Slit
A (or B): Fig. \ref{fig:slits_knots} illustrates the positions of the
putative hidden mass concentrations and our ISAAC long-slit data.

We discuss such analysis below, in two sections: one for ground based
VLT ISAAC spectra and one for the space-based imaging.

\subsection{VLT/ISAAC data}
\label{sec:isaac_analysis}

The ISAAC K-band long-slit spectra hold both kinematic and chemical
abundance information. We discuss the extraction of this information
from the ISAAC spectra after presenting details of our apertures.

\subsubsection{Cluster Apertures}

\label{sec:clust_ap}
To study the properties of the individual star forming clusters at the
centre of M83, individual spectra were extracted from the long-slit
ISAAC spectra. To maximise the SNR and minimise contamination from
background stars and nearby knots, apertures were centred on the peak
flux with widths equal $2\sigma$ (\S\ref{mergeHST+ISAAC}). However,
the poorer seeing-limited resolution of the ground-based ISAAC spectra
sometimes resulted in the blending of clusters and the corresponding
spectra.  This is made clear in Table \ref{tab:cluster_ages} for
clusters that are resolved into multiple components at HST
resolution. Most of the clusters seen in the K-band spectra are
clearly coincident with those detected in H01. However, some are not;
they were too heavily obscured by dust to be detected in the
visible. Table \ref{tab:cluster_ages} identifies clusters already identified in H01.

\begin{center}
  \begin{figure*}
  \includegraphics[width=\textwidth]{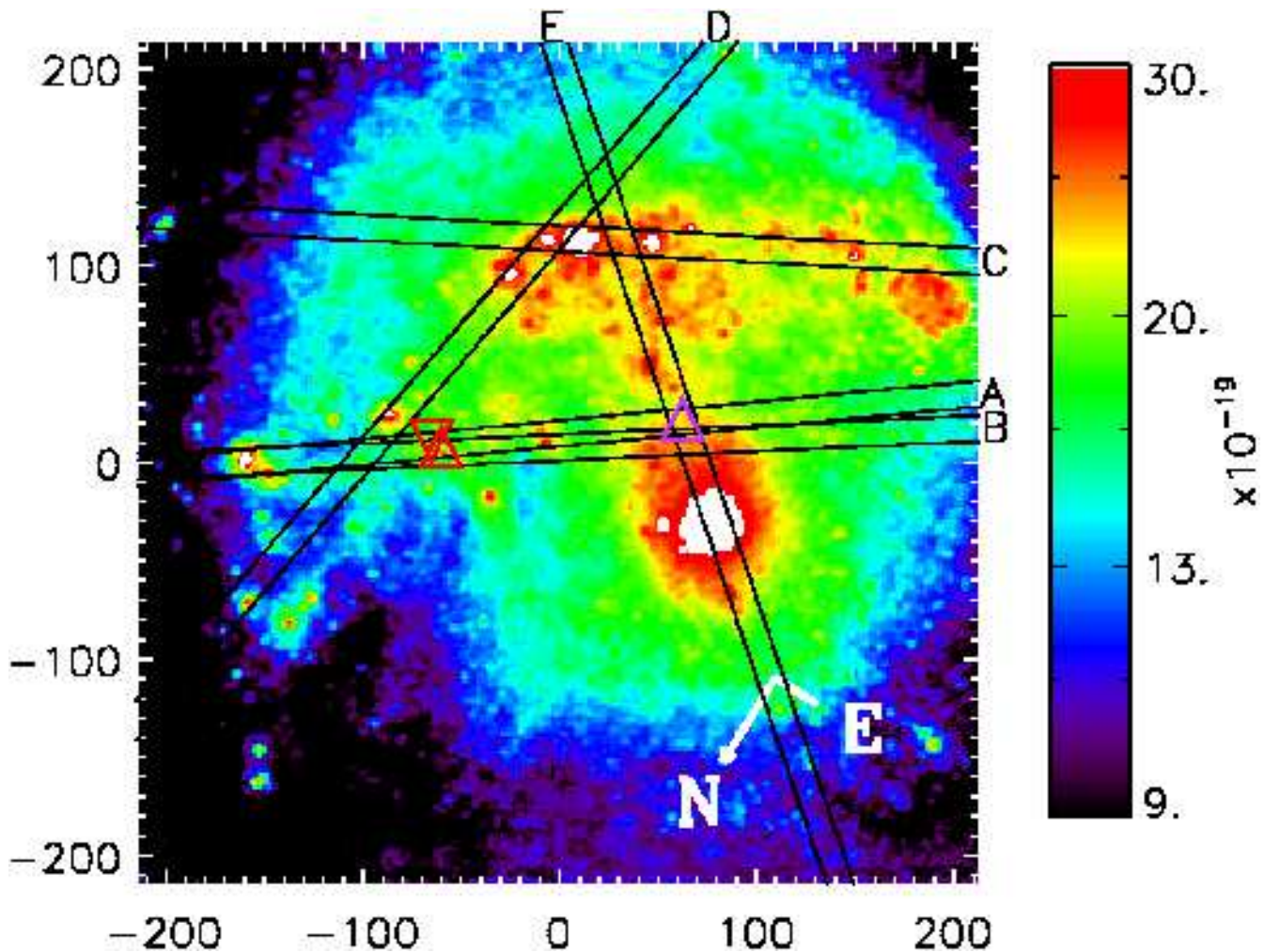}
  \caption{The 5 slit positions overlaid on the F222M NICMOS
  image. Positions of the putative hidden mass concentrations are also
  shown as purple (T00) and red (D06b)
  triangles. For the latter position, two triangles are plotted: the upright triangle gives the position determined from the astrometry while the inverted triangle gives the position estimated directly from the figures of D06b (see \S\ref{sec:errors:astrometry}). Flux values are given in ergs s$^{-1}$ cm$^{-2}$. The position, scaling
  and orientation is the same as Fig. \ref{fig:wn_overlay}.}
  \label{fig:slits_knots}
  \end{figure*}
\end{center}

\subsubsection{Stellar Kinematics}

The stellar kinematics were extracted both for individual clusters and
along the whole slit using a Direct Pixel Fitting technique (based on
that of \citealt{RW92} and described in detail in \citealt{Houghton06}).

The library of stellar templates used to extract the kinematics
contained 20 stars: 6 giants (K1III to M6III) and 14 supergiants (K4I
to M5I). To accurately extract kinematics of a young population (flux
dominated by supergiants), it is essential to have a large number and
range of supergiant templates to minimise template mismatch. This is
particularly important for the CO ro-vibrational absorption after
2.3\micron\ because the depth of the features is strongly dependent on
the spectral type and there is more intrinsic scatter in the
supergiant relation compared to that of giants \citep{KH86,omo93}. The
stellar kinematics were extracted in the range 2.254\micron -
2.355\micron\ (a limit imposed by the stellar templates observed with
the 0\farcs3 slit as described in \S\ref{sec:isaacred} and by the need for good data away from the detector edge).

When extracting stellar kinematics along the entire slit, it was
sometimes necessary to bin-up the data; we chose to bin the spectra to
a minimum SNR of 20 (in the continuum, just before the CO bandhead at 2.295\micron).

\subsubsection{Gas Kinematics}
The neutral gas kinematics of the H$_2$ 2-1 S(1) emission line
\citep[$\lambda_\textrm{vac}22477.17\textrm{\AA}$,][]{H22-1S1} were also
extracted along the slits. This emission does not coincide with the
wavelength range used to extract the stellar kinematics, preventing us
from extracting them simultaneously \citep{Sarzi06}. We instead fitted
a Gaussian profile and linear continuum to the range 2.243\micron\ -
2.258\micron\ (-500\kms\ to +1500\kms\ in velocity; approximately
$\pm$1000\kms\ from the average gas velocity).

\subsubsection{\wco}
We measured the equivalent width of the first CO band ($\mu=2\to0$),
\wco\ (as defined by \citealp{omo93}) from the ISAAC K-band spectra. \wco\
was calculated in the traditional sense,
\begin{equation}
  \textrm{W} = \sum_{i} \Big[ \frac{f_i}{c_i} - 1 \Big] \delta\lambda,
\end{equation}
where $f_i$ is the observed (line and continuum) flux of each bin,
$c_i$ is the (estimated) continuum flux of each bin, $\delta\lambda$
is the width of each bin (in \AA) and $i$ indexes a suitable
wavelength range.

However, \citet{ookm95} showed that the measured \wco\ of a galaxy is
affected by the dispersion of the system. We corrected for this using
the same technique as in \cite{Houghton06}: we artificially
broadened the stellar kinematic templates, fitted a second order
polynomial to the mean result and then applied a correction using the
measured stellar velocity dispersion of the cluster. However, for the
most part, this has negligible effect on \wco\ due to the very low
dispersions of the star clusters. 

\subsection{HST/NICMOS and HST/WFPC2 data}

\subsubsection{Extinction map}
\label{sec:WFPC_NIC_EXT}

Using the homogenised \Ha\ (WFPC2) and \Pa\ (NICMOS) images, we were
able to calculate a deep extinction map for the centre of M83 as well
as for individual clusters. We used a standard Milky Way extinction
curve \citep{Cardelli89}, recombination ratios of \citet{Osterbrock89}
and a ratio of total to selective extinction, R$_\textrm{\scriptsize
V}$, of 3.1 to calculate the extinction \Av.  Thus we find:
\begin{equation}
  \textrm{A}^{gas}_\textrm{\scriptsize V} = 3.1~\textrm{E}(B-V)^{gas} = -1.6
\ln{ \left[ \frac{
  \textrm{F}_{\textrm{\scriptsize H}\alpha} /
  \textrm{F}_{\textrm{\scriptsize Pa}\alpha} } {8.13} \right] }
  \label{eq:av}
\end{equation}
where $\textrm{F}_{\textrm{\scriptsize H}\alpha}$ and
$\textrm{F}_{\textrm{\scriptsize Pa}\alpha}$ are the line fluxes of
the \Ha\ and \Pa\ emission, respectively. The superscript `gas'
identifies that this term has been calculated from the ionised
gas. \citet{Calzetti94} found a substantial difference in optical
depths between the \Ha\ and \Hb\ Balmer emission lines and the stellar
continuum underlying these two Balmer lines (in starburst
galaxies). This was interpreted as a consequence of the hot ionising
stars being associated with more dustier regions than the bulk stellar
population. Our $A^{gas}_V$, determined from the \Ha:\Pa\ decrement,
will suffer a similar effect and we interpret it as the extinction
suffered by the ionised gas, \emph{not} the extinction of the
(continuum producing) stellar population. For brevity, we now drop the
superscript \emph{gas} in \Av\ and \Ebv. 

\begin{center}
  \begin{figure}
  \includegraphics[width=0.5\textwidth]{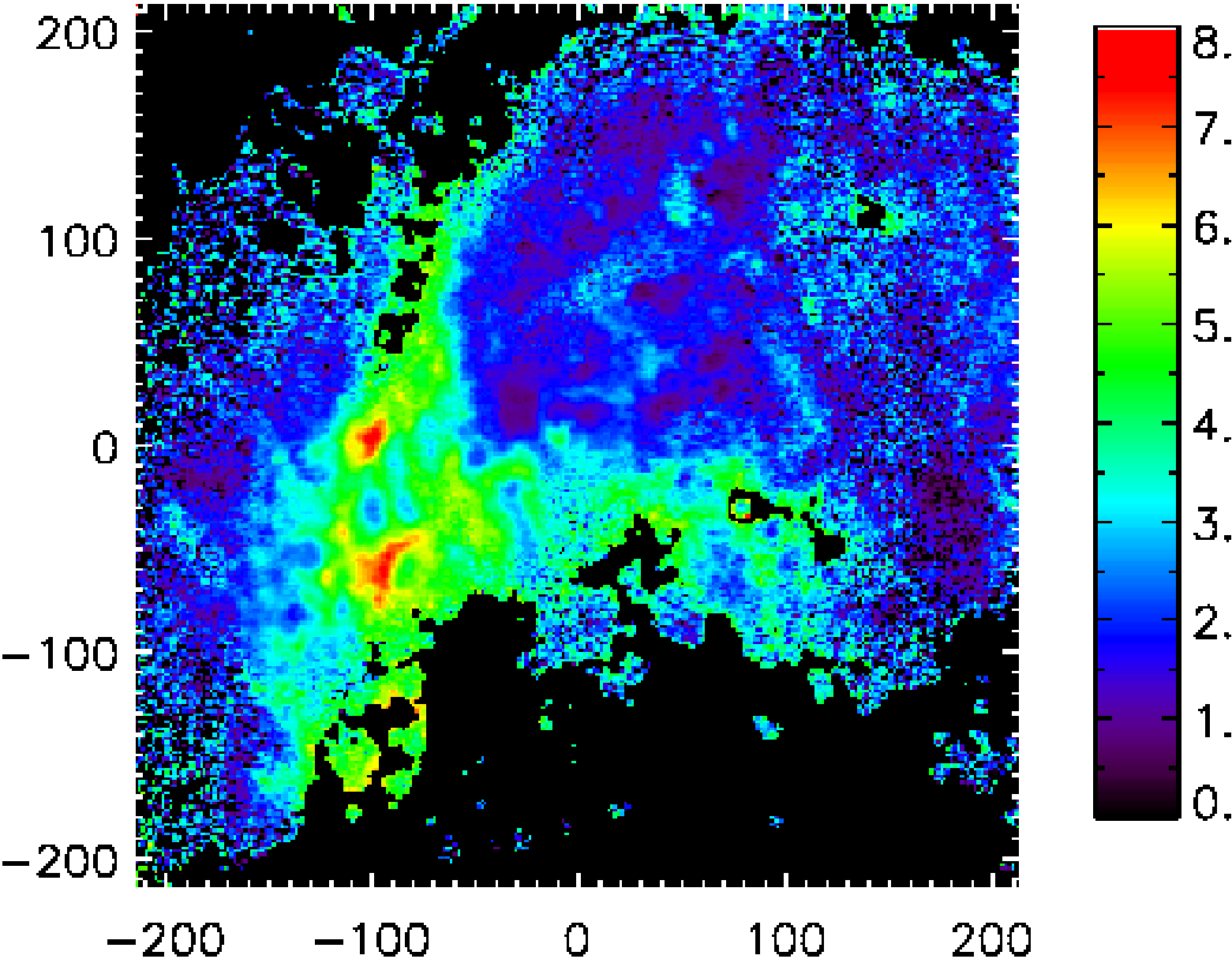}
  \caption{The extinction \Av\ calculated from the \Ha:\Pa\ ratio. The
  position, scaling and orientation is the same as
  Fig. \ref{fig:wn_overlay}. Only pixels
  with $2\sigma$ detections in either the \Ha\ or \Pa\ fluxes are
  shown (pixels not matching this criterion are coloured black). A
  substantial fraction of these pixels have no secure upper bound on
  the extinction value, within the $2\sigma$ errors; almost none have
  a secure upper limit within $3\sigma$. This is a result of the low
  SNR of the NICMOS \Pa\ data.}
    \label{fig:ext_map}
  \end{figure}
\end{center}

Figure \ref{fig:ext_map} illustrates the extinction map. A substantial
fraction of the pixels in the map have no secure upper limit; that is
to say, within $2\sigma$ errors, they are unconstrained. A similar
result evident in Fig. \ref{fig:A_avkin}.
 
\subsubsection{Cluster Apertures}

\label{sec:calzetti}

Apertures with the exact same size and position as those for the ISAAC
data (\S\ref{sec:clust_ap}) were extracted from the synthetic HST
slits (\S\ref{mergeHST+ISAAC}), to calculate the equivalent widths
of the \Pa\ and \Ha\ emission for individual clusters.

As \wha\ and \wpa\ were derived from homogenised narrow band images,
we calculate an approximation to the true equivalent width, namely
\begin{equation}
  \textrm{W} \simeq \Big[ \frac{ F_\lambda - C_\lambda}{C_\lambda} \Delta\lambda
\Big] ,
\end{equation}
where $F_\lambda$ is the observed flux density (line and continuum) of
the narrow band image, $C_\lambda$ is the flux density of the
continuum and $\Delta\lambda$ is the bandwidth of the narrow band
filter.

These indices also require correction. As mentioned previously, when
studying starburst galaxies, \citet{Calzetti94} found different
optical depths for the \Ha\ and \Hb\ Balmer emission lines and the
continuum underlying these two Balmer lines. Consequently, the
observed equivalent widths of ionised gas emission are affected by
this differential extinction between gas and stars and require
correction before comparing to models.

Using the prescription described in \citet{Calzetti97} and
\citet{Calzetti01}, the extinction corrections for \wha\ and \wpa\ were
calculated to be
\begin{equation}
\log{\left[\frac{\textrm{W}(\textrm{H}{\alpha})_o}{\textrm{W}(\textrm{H}{\alpha}
)_i
  }\right]} = -0.4~\textrm{E}(B-V)
\end{equation}
\begin{equation}
\log{\left[\frac{\textrm{W}(\textrm{Pa}{\alpha})_o}{\textrm{W}(\textrm{Pa}{
\alpha})_i
  }\right]} = -0.04~\textrm{E}(B-V)
\end{equation}
where $\textrm{W}_o$ is the observed equivalent width, $\textrm{W}_i$
is the intrinsic width and $\textrm{E}(B-V)$ is calculated from
the \Ha:\Pa\ decrement using a `standard' extinction curve
(\S\ref{sec:WFPC_NIC_EXT}). The errors in the corrected values were
nearly always dominated by the error in $\textrm{E}(B-V)$.

\subsection{Starburst99 Ages}\label{sec:sb99}

SB99 models enable us to age-date individual knots seen in the star
forming region of M83. We use three indices for this purpose: \wco\
direct from the stellar population and \wha\ and \wpa\ from the
gaseous nebula emission (which essentially measure the same quantity,
but differ due to the effects of extinction). \wco\ is defined
according to \citet{omo93} in both SB99 and our measurements. Although
\wco\ is affected by the spectral resolution of the observations, the
low resolution of the SB99 model spectra in the NIR is not an issue
because SB99 calculates \wco\ from expanded versions of the (higher
resolution) \citet{Origlia99} models, which in turn implement the
theoretical equivalent widths from \citet{omo93}.

Although many nebula emission indices are calculated by default in the
SB99 models, \wpa\ is not. Therefore, we follow the prescription of
\citet[and references therein]{LH95} and calculate it directly from
the number of ionising photons (with wavelength $< 912$\AA) and the
model \Pa\ continuum at 1.876\mum\ as a function of time.

\subsubsection{SSP models}\label{sec:sb99:sspmodels}

Simple instantaneous single stellar population (SSP) burst models for
\wpa\ and \wco\ (Fig. \ref{fig:sb99eqws}, dashed lines) were
calculated with a Salpeter IMF ($\alpha=2.35$), a mass range of
1\msun$< M_{\star} < 100$\msun\ and for a metallicity $Z=2Z_{\odot}$
\citep[appropriate for M83]{METAL} using the Geneva tracks
\citep{Geneva}. No \wco\ is expected before 6 Myrs, which is the point
at which the massive stars of a SSP first evolve to the red
super-giant (RSG) phase; this property is relatively insensitive to
variations in $M_\textrm{\scriptsize up}$ between 30 and 100 \msun\
because of the small number of RSGs evolving from progenitor masses
above 30 \msun\ \citep{Origlia99}.  These SSP models were found to be
incompatible with our data for individual clusters: \wpa\ gave
consistently younger ages compared to \wco, by order of a few Myrs.
If the RSG phase was occurring earlier than predicted by the models,
one would expect all \wco\ measurements to be $>15$ as the time period
for which young stars have $0 <$ \wco\ $< 15$ is very small. We
therefore inferred that the variation in \wco\ comes from a
\emph{mixed} population, of stars that have evolved to the RSG and
stars that have not, which in turn implies a mix of ages and a finite
formation timescale for the clusters.

\subsubsection{Mixed population models}\label{sec:sb99:mixedmodels}

SB99 models only provide two cases of the \emph{star formation law}:
an instantaneous burst or constant star formation. We therefore
investigated the effect of different formation scenarios using the
first of these limiting cases. In particular we investigated the
evolution of a short, finite episode of star formation. Similar star
formation laws (albeit for exponential bursts) were investigated in
\citet{FS03}, but using different models and data.

All indices for the mixed population were calculated by convolving SSP
model fluxes (or flux densities) over time with a suitable
kernel. Convolution with a top hat kernel thus mimicked a finite
episode of constant star formation. \wpa\ was calculated by
convolving the number of ionising photons of a SSP and the related
model continuum flux density at 1.876\mum\ with the desired kernel. We
then calculate \wpa\ as before using these new mixed population
values. When calculating \wco, we first converted it into a flux
using the model continuum at 2.29\mum. This wavelength-integrated flux
and the same SSP continuum at 2.29\mum\ were then convolved with the
desired kernel and used to recalculate \wco\ for the mixed population.

\begin{center}
  \begin{figure}
  \includegraphics[width=0.5\textwidth]{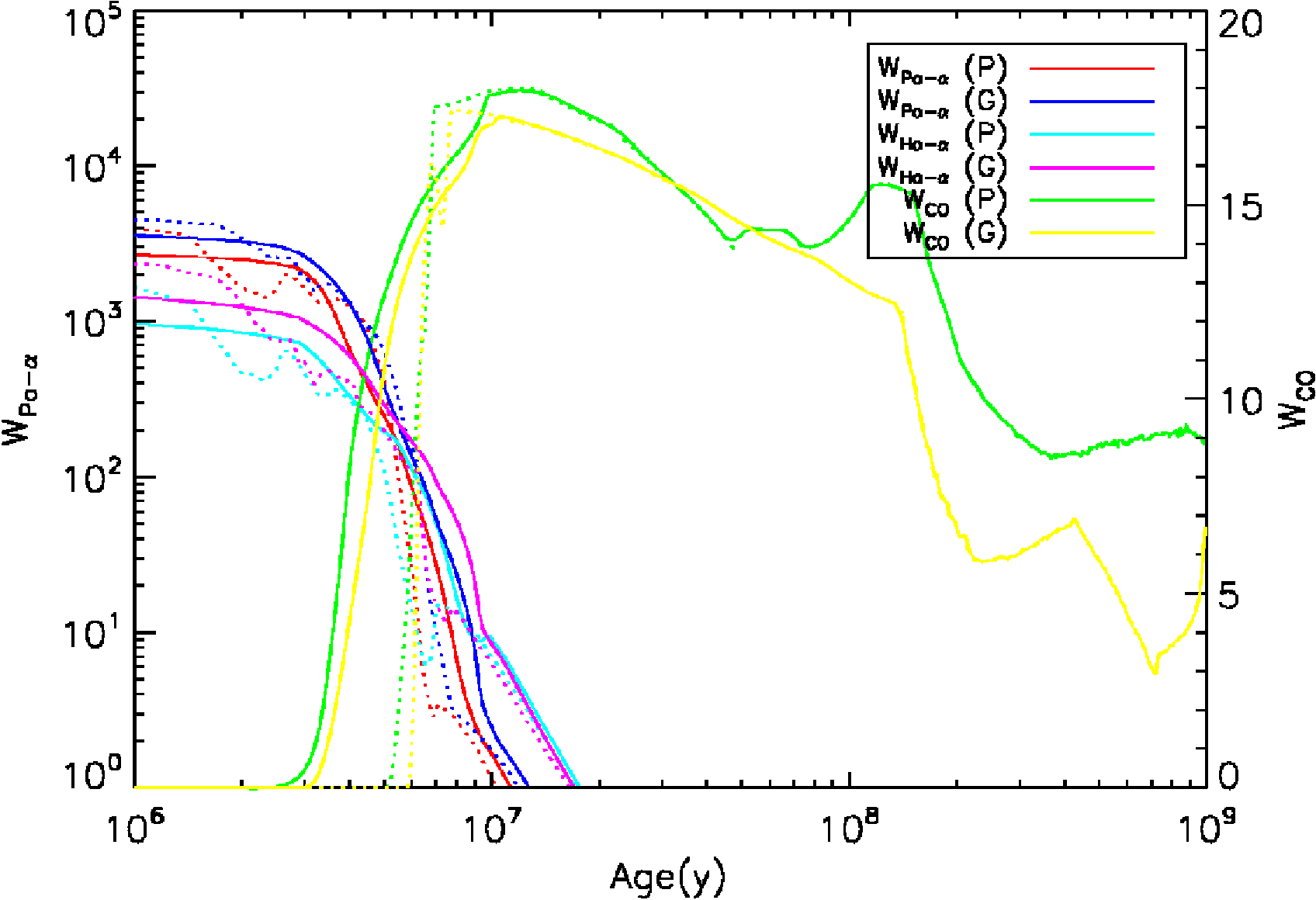}
  \caption{\wpa\ and \wco\ predictions from SB99 with
  $\alpha=2.35$ and M$_{up}=100$\msun. Dotted lines represent the
  instantaneous burst models. Solid lines represent mixed models for a
  finite burst of constant star formation over 6 Myrs. The time
  scale on the plot is relative to the mid-point (mass weighted mean)
  of the burst. Models for both Geneva \citep[G][]{Geneva} and Padova
  \citep[P][]{Padova} tracks are shown, thanks to the recent update to
  SB99 \citep{SB99update}.}
    \label{fig:sb99eqws}
  \end{figure}
\end{center}

Fig. \ref{fig:sb99eqws} shows \wpa\ and \wco\ for an episode of
continuous star formation lasting 6 Myrs (solid lines).  The effect of
a finite duration star formation on \wpa\ is clearly negligible, but
the onset of the \wco\ index is brought forward in time by a few
Myrs. As expected, after around 10 Myrs, the SSP and mixed population
models are indistinguishable as changes in the indices become slower
and less pronounced. The mixing timescale of 6 Myrs is somewhat
arbitrary and was chosen by eye to to best fit the data. Formal errors
are difficult to estimate, although the data appears to be
inconsistent with timescales less than 5 Myrs and larger than 7 Myrs,
giving a 1 Myr margin. 

Strictly speaking, we cannot perfectly reproduce a finite duration of
constant star formation from instantaneous burst models; rather, we
are in fact producing a mixed population via many sequential
instantaneous bursts over a finite duration and assuming that this
closely approximates a short episode of constant star formation. The
time resolution for the mixing (separation between the sequential SSP
models) is 0.1 Myrs. We see no significant change with smaller time
resolutions and so presume that we closely approximate the ideal case
of continuous star formation over a finite time.

Using the same techniques, we investigated the effect of the functional form for the
star formation during the finite episode and found reasonable fits for a finite burst of exponentially increasing star formation (e-folding time $\sim3$Myrs) and a finite burst of exponentially decaying star formation (e-folding time $\sim2$Myrs). Although the finite episode of constant star formation always provided an overall better fit, our data was not able to reliably distinguish between such alternatives, so we chose to adopt the simplest scenario (i.e. a short burst of constant star formation).
We also tested the effect of contaminating
an SSP model with an old (Gyr) population and such a
scenario could not reproduce the observed data trends: the \wco\ estimates were too high for the younger clusters (i.e. those with high \wpa). Finally, we tested multiple SSP bursts separated by the dynamical timescale and half the dynamical timescale, motivated by the work of \citet{Allard06,Sarzi07} and \citet{FB07} and our attempts mostly suffered the previous problem: too much \wco\ at younger ages. However, although many episodic bursts (i.e. more than 2) were unsuccessful at reproducing the observed data trends, models with only two burst were more successful: fine tuning the mass fraction of the latter burst to be 3 -- 4 times the initial burst mass, setting the delay between the bursts to 5 Myrs and diluting the \wco\ with a fraction of the total luminosity to mimic hot dust emission (somewhat simplistic compared to the models of \citealt{FS03}) and replacing the SSP bursts with finite (Myr) bursts went some way to reproducing the data but did not completely resolve the discrepancy. Thus, despite considerable fine tuning of various parameters, multiple bursts did not match the data as well as a constant burst lasting 6~Myrs.

SB99 was recently updated to offer models based on either the Geneva \citep{Geneva} or the Padova tracks \citep{Padova}. Although we illustrate models using both the Geneva and Padova tracks in Fig. \ref{fig:sb99eqws}, we use Geneva tracks to estimate the ages of the clusters in the final analysis (the same as H01); the Padova tracks with a similar metallicity (2.5Z$_\odot$) still match the data and provide a similar age gradient, but with a tendency for younger ages. This is explained by the slightly earlier onset of \wco\ in Fig. \ref{fig:sb99eqws} when using the Padova tracks.

It is not trivial to read the cluster ages from an age-metallicity grid such as Fig. \ref{fig:wpa_v_wco}: if one allows variation in the metallicity, there is a larger range of acceptable ages given the age-metallicity degeneracy. As the random errors in \wpa\ are usually much larger than those of \wco, one could use a fixed metallicity ($2Z\solar$) and take the nearest age along the horizontal but such an approach is unfeasible when the error for \wpa\ is small compared to the horizontal distance to the nearest model track (as is the case for cluster 8). We derived ages from models of a fixed metallicity ($2Z\solar$) using the shortest line from the observation to the $2Z\solar$ model tracks along \wpa\ \emph{and} \wco, effectively ignoring the larger uncertainty in \wpa\ and any variation in metallicity\footnote{Perhaps a better method would be to weight the difference in \wpa\ and \wco\ between the models and observation with the observational errors and then find the closest model point using these scaled distances, but in our case, the rewards of the increased accuracy would be minimal and somewhat futile considering the systematic errors discussed in \S\ref{sec:errors:ages}}. We show the $Z\solar$ models (and how they link with the $2Z\solar$ models) in Fig. \ref{fig:wpa_v_wco} to justify our metallicity assumption: most of the clusters scatter around the $2Z\solar$ tracks.

\section{Errors}
\label{sec:errors}

The estimation of errors in our datasets deserves special mention as
they sometimes required care and are not always calculated
using the standard first-order approach, for good reason.

\subsection{Homogenised HST data}

By homogenising the HST datasets, we introduced correlations between
the pixels and between the errors. Whenever we added pixels, we added
the variances. This would usually cause our errors to be severely
underestimated given that the variances become correlated and would
not add in such a manner. However, careful scrutiny of our method
using Monte-Carlo simulations of noise frames actually indicates that
our final random errors are \emph{over-estimated} by a \emph{minimum}
factor of $\sim 1.7$. The cause is the seeing convolution stage: we
simply convolve our variance arrays in the same manner as the data,
but this only results in obtaining a seeing weighted average variance
for each input pixel; in reality, the true error is reduced much more
than this. Thus, the greater the seeing, the larger our
over-estimation of the errors. We estimate factors of
(1.7, 1.9, 1.9, 2.9) for slits (A, B, C, D), respectively but we do
\emph{not} apply these corrections.

One could question why we degraded the \Ha\ and \Pa\ data to match the
spatial resolution of the ground-based data: we could obtain better
spatially resolved data for these indices. However, then one is faced
with the dilemma of combining non-homogeneous data and which to
trust. The final spatial resolution of our data is thus limited by the seeing, but it is worth recalling the resolution of the study by H01, which is only a factor of two better. Although the diffraction limited FWHM of HST at 656nm is 0\farcs056, the pixel sampling of WFPC2 is 0\farcs04554. However, H01 further convolved the data with a Gaussian of $\sigma=2$ pixels~(FWHM = 0\farcs21) to create sufficiently uniform cluster profiles that could be well fitted by a single PSF model. This brought the final spatial resolution of the HST data to 0\farcs22. The resolution of our best data is 0\farcs4 --- only a factor of two larger. Furthermore, H01 calculated \wha\ within 0\farcs56 and 1\farcs34 diameter
apertures (12pc and 29pc at 4.5Mpc) and found little difference between
them for most clusters. Our apertures are generally within this range:
they are rectangular with heights equal to the slit width (0\farcs6)
and widths equal to (0\farcs3, 0\farcs4, 0\farcs4, 1\farcs1) for slits
(A, B, C, D), as determined by the seeing. We choose to use the higher
resolution of the space-based data only to flag potentially
contaminated (unresolved) data in Table \ref{tab:cluster_ages}. 
Thus we measure luminosity weighted
ages for any contaminated clusters which may be particularly true for
clusters 5, 8 and 12 in Table \ref{tab:cluster_ages}.

\subsection{\wco}\label{sec:errors:wco}
If we calculate the random error in \wco\ from error in the flux
values alone, we find that it is negligible compared to the error
introduced by uncertainly in the velocity and dispersion. Therefore,
we use $\pm1\sigma$ random errors in the velocity and velocity
dispersion measurements of each cluster to calculate the systematic
error in \wco. We then use whichever is the larger. As a result, the
error in \wco\ is often asymmetric.

\subsection{\Ha}\label{sec:errors:wha}
As discussed in \S\ref{sec:wfpc2_images}, the \Ha\ flux is subject
to an uncertainty in the gas velocity due to the systematic velocity
of M83 placing the emission on the edge of the F656N bandpass.  We
therefore appoint \emph{minimum} errors of 10\%\ to all \Ha\ fluxes
(often dominating the error in \wha\ for the clusters) and in
calculating the extinction estimate of Fig. \ref{fig:A_avkin}. This
10\%\ systematic error corrects for velocity range of 430\kms\ to
560\kms. Because this error is systematic, it does not change with
binning or aperture size and is applied as a last step before
analysis.

\subsection{\Av, \Ebv}
\label{sec:errors_av}

Extinction estimates using a HII decrement are often tricky due to the
uncertain penetration depth of the emission lines in question:
estimating the extinction in dusty regions using the \Hb:\Ha\
decrement invariably leads to a lower estimate than comparing the
\Ha:\Pa\ decrement as the effective penetration is much greater with
the \Pa\ line. We assume that our \Ha:\Pa\ decrement is not limited in
this way for the range considered here $(0 <$\Av$< 8)$. However,
estimating the random error is still non-trivial.

A standard first-order approach gives that the error in
Eq. \ref{eq:av} is $\sigma_{A_v} =
-1.6({\textrm{F}_{\textrm{\scriptsize
Pa}\alpha}}/{\textrm{F}_{\textrm{\scriptsize H}\alpha}})$. However,
this is only valid for \emph{small} errors in
${\textrm{F}_{\textrm{\scriptsize Pa}\alpha}}$ and
${\textrm{F}_{\textrm{\scriptsize H}\alpha}}$: the errors in \Pa\ data
are large (sometimes $\gg10$\%) and the approximation breaks down. We
therefore calculate the error explicitly by substitution. Due to the
non-linear nature of Eq. \ref{eq:av}, there is considerable difference
between the $1\sigma$ and $2\sigma$ confidence limits. We therefore
opt to quote $2\sigma$ values because $1\sigma$ values are
significantly smaller and lead to a false impression of accuracy. The
errors in the corrected \wha\ and \wpa\ of Fig. \ref{fig:wpa_v_wco}(c)
and (d), dominated by the error in \Av\ were calculated
using this $2\sigma$ error.

\subsection{Cluster Ages}\label{sec:errors:ages}

We wish to quote an error for the ages of the clusters, but this is not trivial. Here, we list the reasons for the difficulty and explain how the age errors in Table \ref{tab:cluster_ages} are determined.

The cluster ages are calculated from the NIR indices \wpa\ and \wco. The NIR indices (particularly \wpa) suffer less from extinction and are therefore more accurate in terms of random error and also any systematic error associated with the extinction correction (e.g. the extinction law, the unknown penetration depth of our \Ha\ and \Pa\ data) because the correction is smaller. \wco\ is corrected for velocity dispersion effects but this correction is very small and does not affect the derived ages. 
However, young clusters are expected to be dusty and if heated by UV radiation, this dust
re-emits, mainly in the mid-IR but also in the NIR which could
diminish the \wco\ index by artificially increasing the continuum
level and thus affect our derived ages and the mixing timescale of \S\ref{sec:sb99:mixedmodels}; however, this effect is usually very weak and we do not attempt to correct our data for this effect, unlike
the study by \citet{FS03}.

The choice of evolutionary tracks also plays a role: we quote ages using the Geneva tracks, but the Padova tracks also fit the data albeit with systematically lower ages of around 0.5 Myrs for the youngest clusters (4.5~Myrs old) and 1.5 Myrs for the oldest clusters in the arc (8~Myrs old). This difference leads us to quote a systematic uncertainty in our age estimates of order $\pm1$Myr.

We also vary the mixing timescale of the models to match the cluster data (\S\ref{sec:sb99:mixedmodels}). The mixing timescale was chosen by eye to fit the
observations, which were inconsistent with mixing timescales less than
5 Myrs or larger than 7 Myrs.
Error in this parameter will manifest itself as a systematic error in the ages, particularly for the youngest clusters as this is where the change is greatest. We also note that without
applying the Calzetti et al. extinction law, the data of
Fig. \ref{fig:wpa_v_wco}(a) is best fit by a burst duration of 5.5
Myrs. Thus, given that the mixing timescale is estimated to be accurate to a Myr, we estimate the youngest clusters could also suffer another systematic error of order $\pm1$Myrs.

We are also somewhat limited by our ground-based spatial
resolution. Because we homogenise all our data, this affects \wha,
\wpa\ and \wco\ and the derived ages. If two clusters are unresolved, we expect to measure a
luminosity weighted \wha, \wpa\ and \wco. But whereas younger
clusters ($< 5$ Myrs) will have higher \Ha\ and \Pa\ emission (after
correcting for extinction), the K-band luminosity and the \wco\ is
dominated by older (10 Myr) clusters. Hence if the two unresolved
clusters are different in age, we obtain conflicting ages from the
different indices. However, one expects neighbouring clusters to have
similar ages (given the results of H01), so this effect is probably small. 
Nevertheless, it is possible to identify contamination from another cluster or from the bulge stars (for highly extinct, dim clusters). The luminosity weighted stellar velocity dispersion of low mass star clusters is very low ($<20$\kms), so if we measure them to be high ($>30$\kms), the K-band light is likely not dominated by a single
cluster. Consequently, when a cluster has a
large velocity dispersion ($>30$\kms), it is noted in Table
\ref{tab:cluster_ages}. Furthermore, when extracting the kinematics of each cluster, we fit a linear combination of stellar templates. As discussed in \S\ref{sec:isaacred}, we have a large number of giant and supergiant templates to fit the spectra and the best-fit template can be used to indicate the different stellar populations in the spectrum, \emph{and the relative fractions}. Therefore, in Table \ref{tab:cluster_ages} we give the fraction of giant and supergiant stars used to create the best-fit kinematic template for each cluster. The giant/supergiant fraction clearly correlates with the velocity dispersions and we can identify which cluster measurements are contaminated, and to what degree. What effect the fraction of contamination has on the derived ages is not always obvious, but is likely to lead to bias towards the age of the contaminant; thus we quote the ages (from H01) of any contaminating clusters resolved in HST images in Table \ref{tab:cluster_ages} so that the reader can compare the measured age of the cluster with the measured age of the contaminant(s).

Finally, we have to match the observed cluster data with the models. Given the age-metallicity degeneracy, this is not trivial and we discuss our method in \S\ref{sec:sb99:mixedmodels}. The proximity of our observations to the nearest point on the $2Z\solar$ model track is an indication of how well the model fits the data, but it is also an indication of the stochastic nature of star formation: given a continuous IMF and only a relatively small quantity of gas, the stars formed will not be uniformly distributed over the entire mass range but will be scattered at random positions weighted by the IMF (especially true for the red supergiants). Thus we expect an intrinsic scatter in the number of red supergiants per cluster given a known mass and age, which leads to an intrinsic scatter in the observable quantities \wpa\ and \wco\ --- they will never agree perfectly with the models. Given that H01 estimate most of the clusters to have masses around $10^4$--$10^5$M\solar, we must consider these stochastic processes. For the majority of cases, the agreement between data and models is very good which leads us to estimate a \emph{random} error of 0.5 Myrs for most of the clusters. However, there are some clusters which are more difficult, such as 8 and 10; for these clusters we do not know if error (systematic or random) or nature has resulted in the data falling far from the models. Somewhat arbitrarily, we assign random errors of $\pm1$Myr to these clusters. For clusters that do not match the model predictions for both \wpa\ and \wco\ (clusters 14 and 15) we choose to trust the \wco\ index (see \S\ref{sec:res:cluster_ages}) but assign asymmetric errors to cover the range of all possibilities. For clusters that are do not match the models and are assumed to be very old ($>10$Myrs) we do not quote an error, only a lower limit.

\subsection{Astrometry}\label{sec:errors:astrometry}
Fig. \ref{fig:slits_knots} shows the positions of the putative hidden
mass concentrations relative to the NICMOS data and the ISAAC
slits. 

The slits were located on the NICMOS image as described in
\S\ref{mergeHST+ISAAC} and were \emph{not} positioned using the
astrometry provided with the ISAAC and NICMOS data. Formal
uncertainties are difficult to calculate although by trial and error,
we estimate the error in $x$ and $y$ for Slit A to be at around
0\farcs1 (comparable to a NICMOS pixel) and the error in the position
angle to be at most $1\deg$. However, these values vary for each slit:
with poorer seeing, the ISAAC data is smoother and the errors in the
slit positions increase slightly. 

The pipeline reduced WFPC2 and NICMOS images come with world
coordinate systems. However, the two systems were not comparable: the
cluster positions of H01 were correctly centred on the WFPC2 images
(as to be expected, for they were derived from the same data) but they
were offset when overlaid on the raw NICMOS images (0\farcs54 to the
SE as determined from the position of the nucleus in the
raw F814W image and the raw F222M image). We chose the path of least
resistance and adopted the WFPC2 astrometry. To calculate a world
coordinate system for our homogenised data, we used our precise
knowledge of the pixel size of the homogenised HST data (square pixels
with sides 0\farcs0752963 in length) and varied the absolute reference
to best align the positions of the clusters from H01 on the data. We
assumed no error in the position angle of the WFPC2 and NICMOS data as
defined by the ORIENTAT header keyword. Comparing our astrometry, we
find the nucleus at (13h37m00.91s, -29d51m55.7s) whereas
D06b find it at (13h37m00.95s, -29d51m55.5s): an offset of
0\farcs49 to the NE.

With no agreement between the WFPC2, NICMOS or D06b
astrometry, we retain that of the WFPC2 images.  However, when
plotting the position of the hidden mass concentration given in
D06b we apply a shift of 0\farcs49 to the SW - the same
shift which would align the position of the nucleus given
by D06b to the position of the nucleus in our
world coordinate system (WCS). Quite by chance, more than by design, this
places the position of the putative mass concentration securely in the
middle of Slit A. However, D06b state a $2\sigma$ error of
0\farcs7 in their coordinates for this position; our slit width is
0\farcs6. Thus, it is reasonable to argue that the absolute centre of
the proposed mass concentration could still lie outside our slit
position; however it is unreasonable to argue that we wouldn't detect
its gravitational influence in the stellar kinematics given that the
gas kinematics appear kinematically disturbed over many arcseconds in
data of D06. For completeness, in Fig. \ref{fig:slits_knots} we also plot the position of the D06 interloper using their images as a reference, rather than their coordinates; this second location still overlaps that of Slit A but is approximately 0\farcs38 (8pc) west of the formal position in our WCS.

The position of the mass concentration proposed by T00
was considerably easier to locate: by using the position of the second
dispersion peak in the original data and then locating the position of
this slit on the homogenised HST data using the same techniques as for
the other slits (\S\ref{mergeHST+ISAAC}), we were able to locate
the second dispersion peak on the homogenised HST data without
resorting to any WCS. By design, Slit A overlays
this same position. We assumed that the second dispersion peak was
centred along the slit width, which could be open to
speculation. However, as Slit A is roughly perpendicular to the
original slit of T00 (Slit E in
Fig. \ref{fig:slits_knots}), we would adequately cover any positional
error perpendicular to the original slit.


\section{Results}
\label{sec:results}

We present our findings with regard to the cluster ages and the
putative hidden mass concentrations, in two appropriate sections.

\subsection{Cluster Ages}\label{sec:res:cluster_ages}

\begin{figure*}
  \centering
  \begin{minipage}[c]{0.5\textwidth}
    \centering
    (a)
  \end{minipage}%
  \begin{minipage}[c]{0.5\textwidth}
    \centering
    (b)
  \end{minipage}
  \begin{minipage}[c]{0.5\textwidth}
    \includegraphics[width=\textwidth]{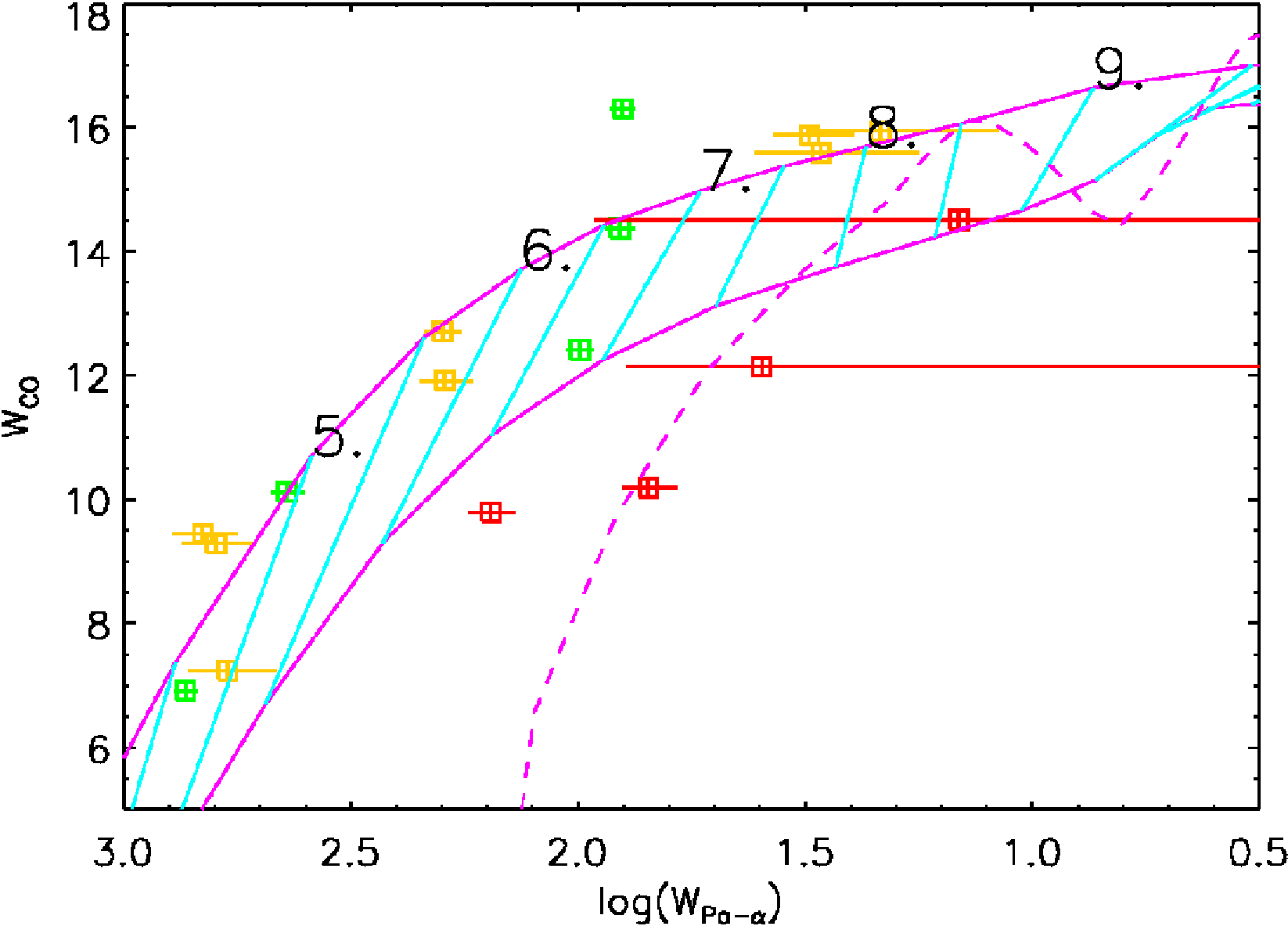}
  \end{minipage}%
  \begin{minipage}[c]{0.5\textwidth}
    \includegraphics[width=\textwidth]{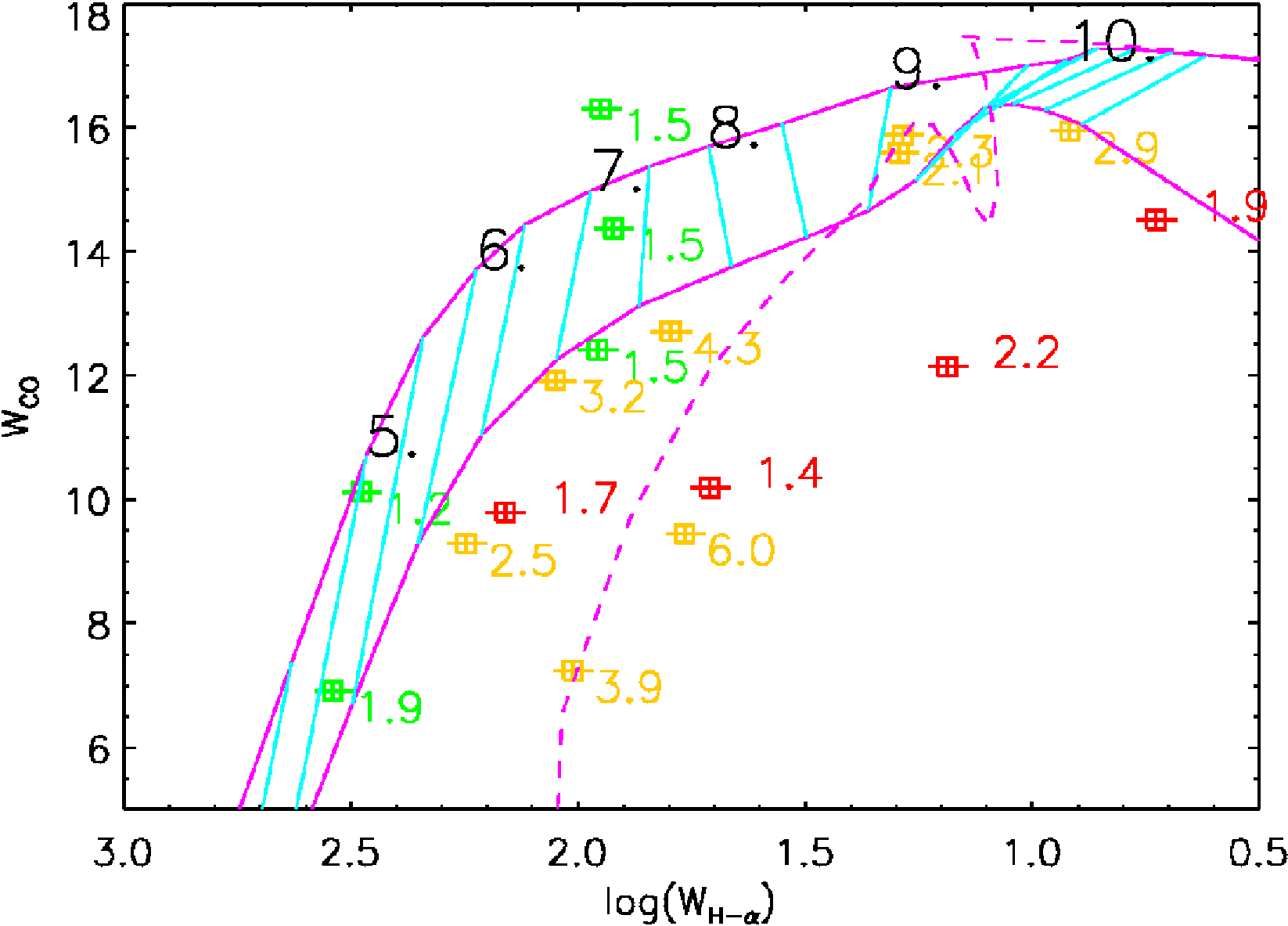}
  \end{minipage}
  \begin{minipage}[c]{0.5\textwidth}
    \centering

    \phn

    (c)

  \end{minipage}%
  \begin{minipage}[c]{0.5\textwidth}
    \centering

    \phn

    (d)

  \end{minipage}
  \begin{minipage}[c]{0.5\textwidth}
    \includegraphics[width=\textwidth]{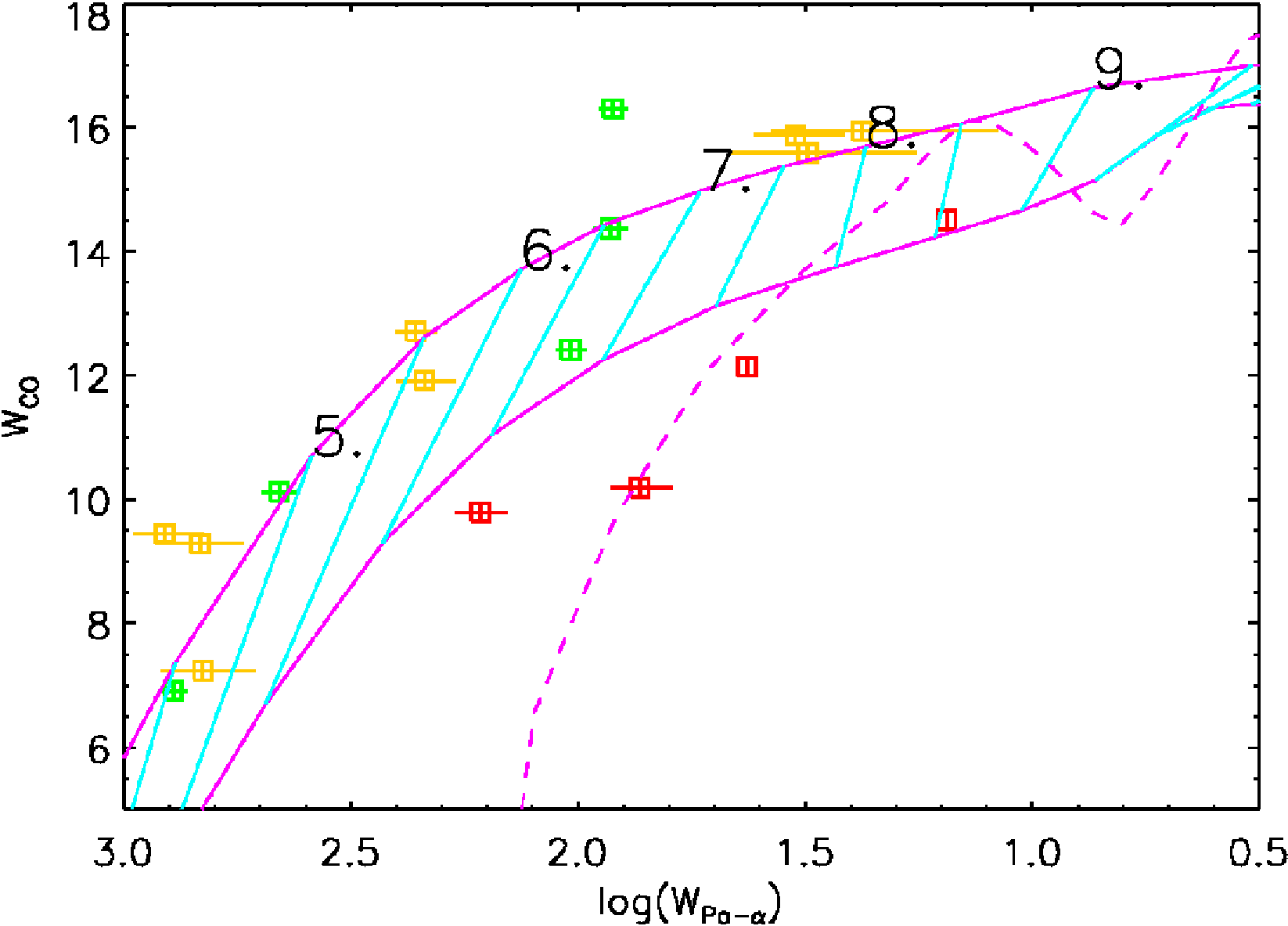}
  \end{minipage}%
  \begin{minipage}[c]{0.5\textwidth}
    \includegraphics[width=\textwidth]{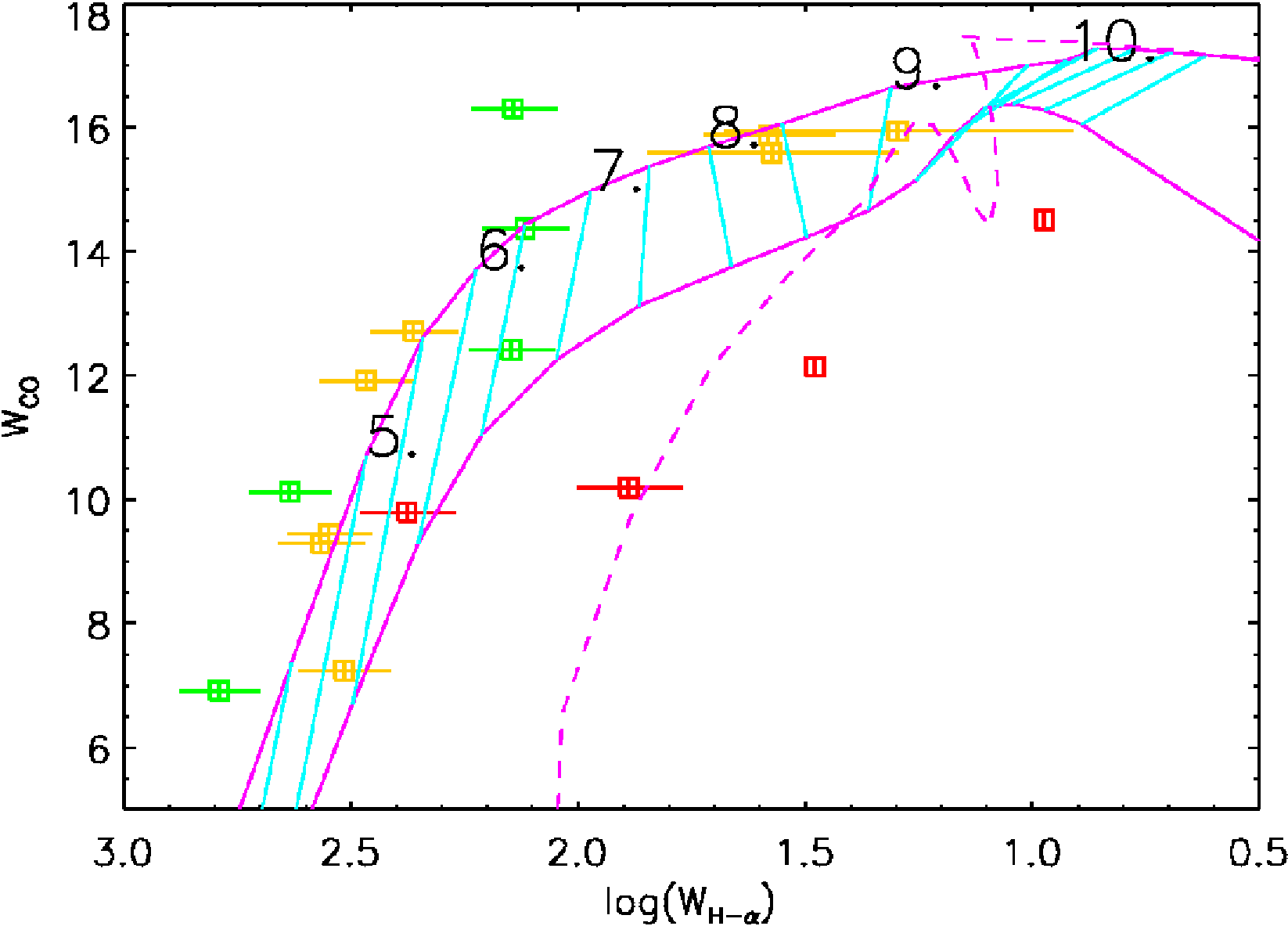}
  \end{minipage}
  \caption{The SB99 models for the mixed population of Fig
  \ref{fig:sb99eqws} with data points for individual star
  clusters. Unlike Fig. \ref{fig:sb99eqws}, two metallicities are
  shown: the top (magenta) curve shows models for Z=2Z$_\odot$ while
  the lower (magenta) curve shows models for Z=Z$_\odot$; the vertical
  blue lines link points of constant age and are spaced in 0.5 Myr
  intervals. Absolute ages are shown in Myrs on the upper (2Z$_\odot$)
  curve. The dashed (magenta) line gives the model predictions of the
  unmixed SSP (2Z$_\odot$). The data points represent measurements for
  individual clusters: green points indicate low extinction clusters
  consistent with ages $<10$ Myrs; yellow points indicate high
  extinction clusters consistent with ages $<10$ Myrs; red points
  indicate clusters that are inconsistent with ages $<10$ Myrs. The
  upper two panels, (a) and (b), show the extincted data; the lower
  two panels show the same plots with \wha\ and \wpa\ corrected for
  differential reddening using the Calzetti prescription
  (\S\ref{sec:calzetti}). Panel (b) also quotes \Av for individual
  clusters. Horizontal error bars in the lower (extinction corrected)
  panels, (c) and (d), are often dominated by the uncertainty in the
  extinction: $2\sigma$ errors are shown for all clusters except for
  two red data points: the extinction, and therefore the corrected
  \wpa\ and \wha, are unconstrained within the $2\sigma$ errors, due to
  low \Ha\ and \Pa\ flux.}
    \label{fig:wpa_v_wco}
\end{figure*}

Fig. \ref{fig:wpa_v_wco}(a) shows the mixed model predictions for the
evolution of \wco\ and \wpa\ for solar (Z$_\odot$) and twice-solar
(2Z$_\odot$) metallicity with data points for individual clusters
over plotted. The same models and clusters are plotted in
Fig. \ref{fig:wpa_v_wco}(b) for \wco\ and \wha. As \wha\ and \wpa\ are
essentially measuring the same quantity (the number of ionising
photons), the main difference between the two figures is the effect of
extinction. 

Green points indicate less extinct clusters (\Av$<2$), that roughly
match the 2Z$_{\odot}$ mixed population models for \wpa\, \wha\ and
\wco, whereas the orange points indicate more extinct clusters
(\Av$>2$), that tend only to match the model predictions for \wpa\ and
\wco\ and lie below the predictions for \wha. At first, \wha\ and
\wpa\ therefore appear to be uncorrelated in
Fig. \ref{fig:wpa_v_wco}(a,b) but after applying Calzetti's
differential extinction, the correlation is obvious in
Fig. \ref{fig:wpa_v_wco}(c,d). The duration of the star formation
episode (6 Myrs) was therefore chosen to fit both the (corrected)
\wha-\wco\ and \wpa-\wco\ data in Figs. \ref{fig:wpa_v_wco}(c,d). 

Red points indicate clusters that do not match model predictions in
the (corrected) \wha-\wco\ and the \wpa-\wco\ plots. Although only
models with ages $< 10$ Myrs are given in Fig. \ref{fig:wpa_v_wco},
these points are not consistent with older populations either, due to
the larger values of \wha\ and \wpa. However, two of these clusters
(16 and 17 in Table \ref{tab:cluster_ages}) have very low \wha\ and
\wpa\ and it is therefore important to consider non-photoionised gas
emission (caused primarily by shocks and their precursors, from
supernovae and massive stellar winds but also turbulent mixing layers
and changes in gas temperature as explained by
\citealt{Calzetti04}). Although these two clusters are not in the
non-photoionised regions of M83 highlighted by \citet{Calzetti04}, one
cannot discount the possibility of non-photoionised emission. 
As they are both quite distinct from
the main star forming arc, we conclude that the \Ha\ and \Pa\ emission
is non-photoionised and thus \wco\ reflects a much older population,
greater than 10 Myrs.
The other two red points (14, and 15 in Table \ref{tab:cluster_ages})
are somewhat puzzling (H01 also found the photometry of these clusters
to be anomalous): either the nebula emission equivalent widths or the \wco, or both, are low
compared to the models. Comparing the ages predicted by each index, we
noticed that the ages predicted by \wco\ were in good agreement with
the ages of nearby clusters, but the ages predicted by \wha\ and \wpa\
were not. Therefore we tentatively assign them ages based on \wco\
measurements alone (although cluster 15 is the only cluster entirely consistent with the instantaneous SSP models, it makes little difference to the derived age).

It should also be noted that clusters 5 and 12 are in a particularly
densely populated region and their ages are likely more representative
of the local average (of those clusters listed as contaminants in
Table \ref{tab:cluster_ages}) because of significant
contamination. Similarly, cluster 9 is probably heavily contaminated
by cluster 13, although we do register significant differences in
\wco\ for the two.

\begin{figure*}
    \centering
    \includegraphics[width=0.9\textwidth]{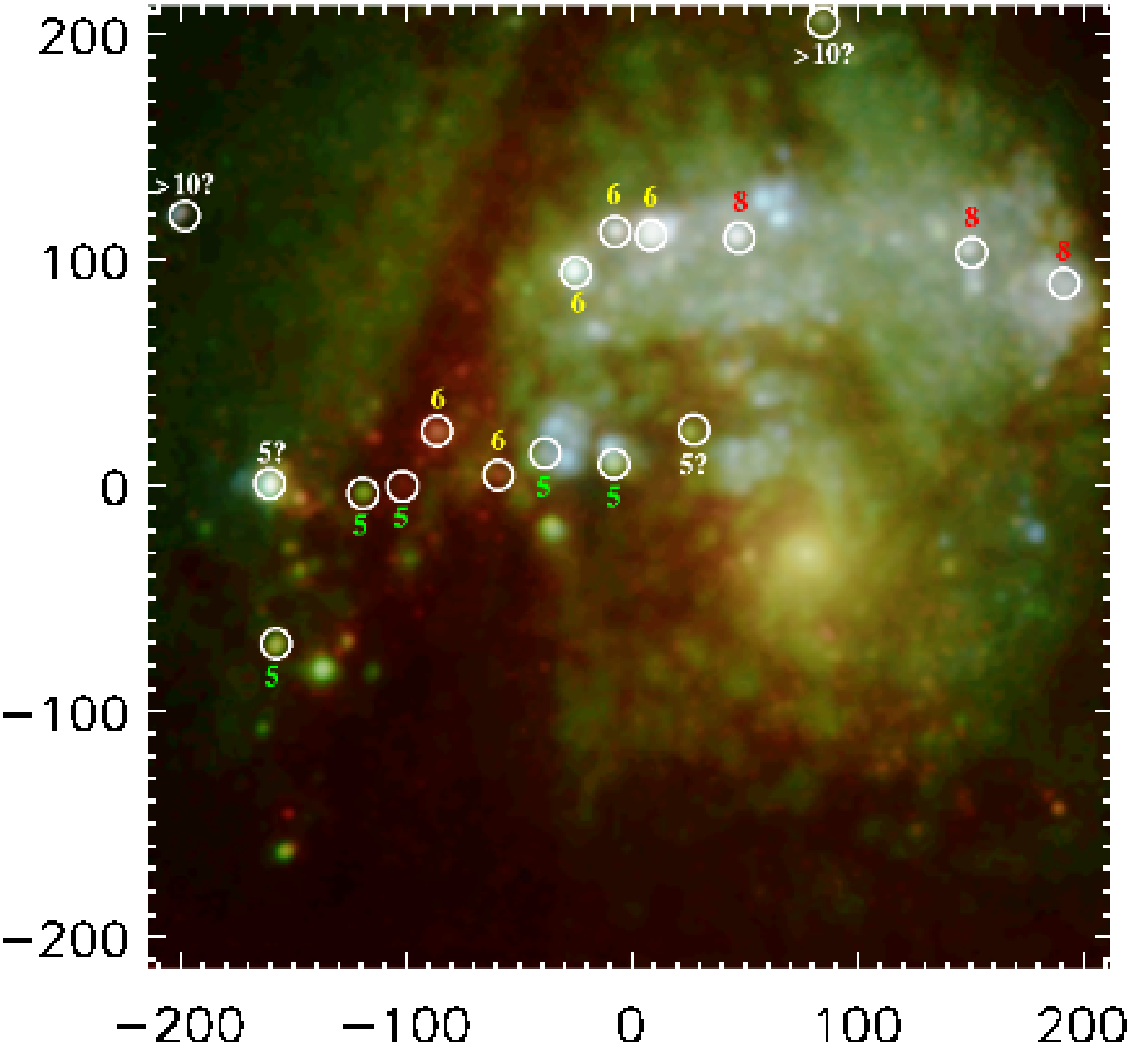}
    \caption{A false-colour image of the centre of M83 made with
    homogenised NICMOS and WFPC2 data. The position, scaling and
    orientation is the same as Fig. \ref{fig:wn_overlay}. 
    NICMOS F222M is red, WFPC2 F814W is green and
    WFPC2 F300W is blue. A non-linear grey-scale has been employed to
    enhance low luminosity (heavily extinct) clusters. Note that the
    circumnuclear arc can be seen to extend into and along the dust
    lane. Cluster ages (derived from Fig. \ref{fig:wpa_v_wco} and
    rounded to the nearest Myr) are over plotted: colours of the ages
    aid the eye in distinguishing the age gradient along the arc;
    green represents 5 Myrs, yellow represents 6 Myrs, red represents
    8 Myrs while white represent uncertain ages. The resolution of this image is set by the NICMOS instrument and we estimate a FWHM of 0\farcs15.}
    \label{fig:cluster_ages}
\end{figure*}

Fig. \ref{fig:cluster_ages} illustrates the positions of the clusters with their ages (rounded to the nearest Myr) from Fig. \ref{fig:wpa_v_wco}(c), while Table \ref{tab:cluster_ages} quotes them to the nearest 0.5 Myrs. Although the ages from Fig. \ref{fig:wpa_v_wco}(c) and Fig. \ref{fig:wpa_v_wco}(d) agree remarkably well for each cluster, we use the age derived from the NIR data (\wpa) because it is less effected by extinction and is therefore more accurate (as discussed in \S\ref{sec:errors:ages}).

\subsection{Hidden Mass concentrations}
\label{results:hmc}

\begin{figure}
    \centering
    \begin{minipage}[c]{0.5\textwidth}
      \includegraphics[width=\textwidth]{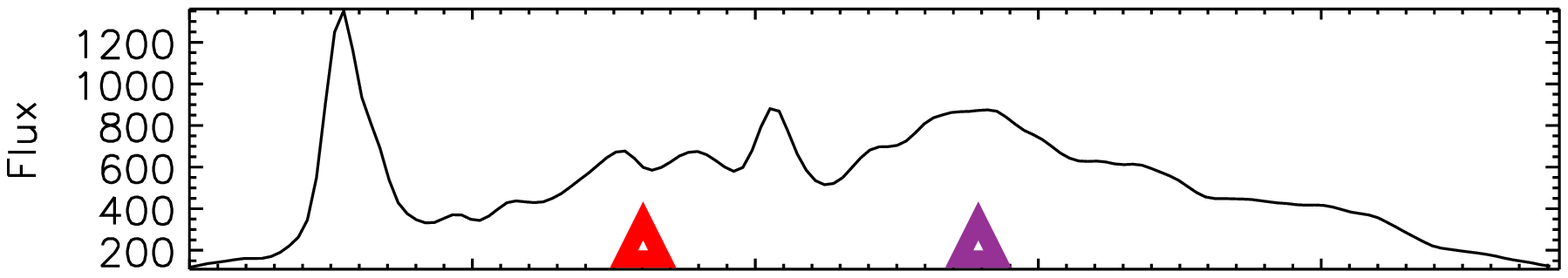}
    \end{minipage}
    \begin{minipage}[c]{0.5\textwidth} 
      \includegraphics[width=\textwidth]{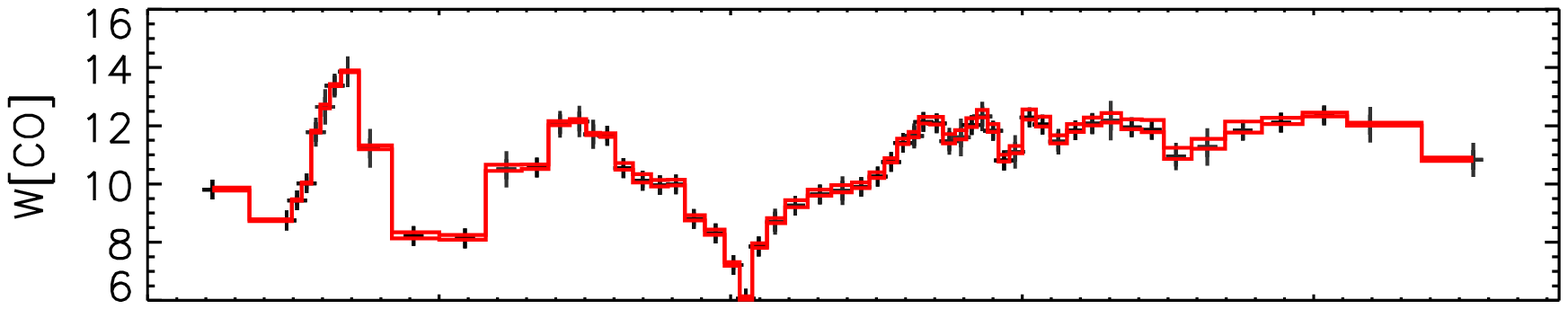}
    \end{minipage}
    \begin{minipage}[c]{0.5\textwidth} 
      \includegraphics[width=\textwidth]{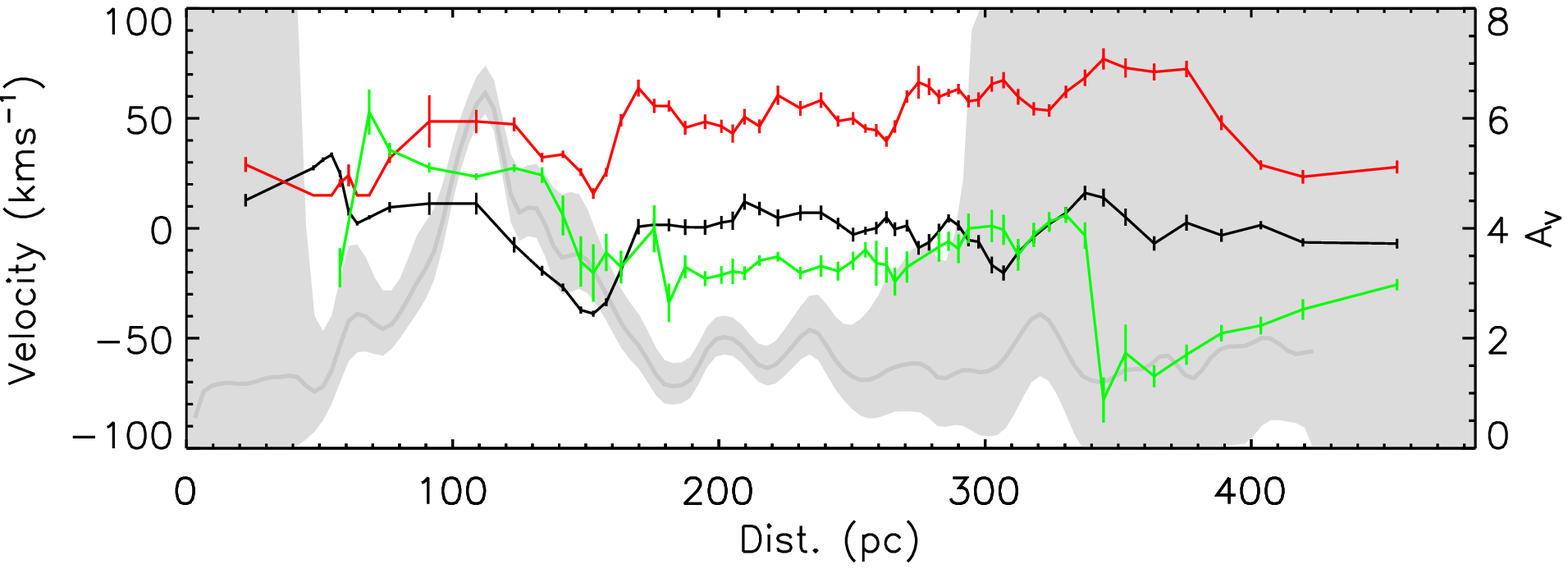}
    \end{minipage}

    \caption{\emph{Upper panel:} the flux profile of slit A. Positions
    of the putative hidden mass concentrations are also shown as
    purple (T00) and red (D06b)
    triangles. \emph{Middle panel:} The \wco\ measured in each bin
    along slit A: red lines indicate maximum systematic errors from
    velocity and dispersion uncertainty; the error bars on the points
    correspond to random error. \emph{Lower panel:} The Kinematics
    along slit A: the black, red and green lines illustrate the
    stellar velocity (relative to galaxy's mean stellar velocity), stellar velocity dispersion
    and the H$_2$ gas velocity (also relative to galaxy's mean stellar velocity),
    respectively; the grey line shows the extinction, \Av\ calculated
    from the \Ha\ and \Pa\ emission and the lighter grey shading
    illustrates the $2\sigma$ error. \emph{Both panels:} The PA of the
    slit is 126.95\Deg, with the northern side on the left.}
    \label{fig:A_avkin}  
\end{figure}

Fig. \ref{fig:slits_knots} illustrates the positions of the putative
hidden mass concentrations of D06 and T00 and our long-slit data. 
Both slits A and B
show identical kinematic features at the location of D06b,
therefore for brevity we show only the the kinematics along Slit A in
Fig. \ref{fig:A_avkin}.
 
A sharp gradient in the H$_2$ gas velocity \Vg\ is found in the same
vicinity that D06 report a gradient in the $\textrm{Pa}\beta$ gas
velocity (140pc along slit A) and both gradients are in the same
direction. The gradient in H$_2$ is sustained (i.e. \Vg\ is roughly
constant for 80pc to the north). There appears to be a gradient in the
opposite direction at 60pc along the slit, although the line flux is
low, the kinematics of Slit B disagree and there are luminous star
clusters at this location which might bias the observations.

The stellar velocity \Vs\ and stellar dispersion \Ss\ profiles both
show synchronous drops in the vicinity of the D06 mass concentration
(160pc along slit A), but these changes are not sustained and \Vs\ and
\Ss\ both recover, although the recovery is significantly more gradual
on the north side. 

All these features occur approximately at the edge of the
north-east dust lane, clearly seen by plotting the extinction (\Av)
along the slit in Fig. \ref{fig:A_avkin}.

\begin{figure}
    \centering
      \includegraphics[width=0.5\textwidth]{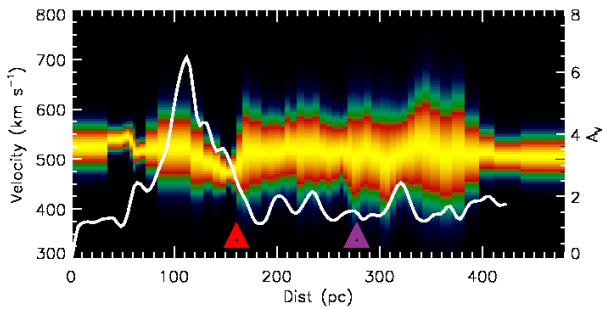}
    \caption{A position velocity diagram of the stellar kinematics
    from Fig. \ref{fig:A_avkin}. Where it was necessary to bin the
    data to extract a VP, they have been stretched to represent the
    space over which the data was binned. The extinction profile \Av\
    of Fig. \ref{fig:A_avkin} is over plotted in white. Positions of
    the putative hidden mass concentrations are also shown as purple
    (T00) and red (D06b) triangles. }
    \label{fig:p-v}
\end{figure}

In Fig. \ref{fig:p-v}, we show a position--velocity diagram for the
stellar kinematics. The best-fit Gaussian line-of-sight velocity
profiles (VPs) are shown at each position along the slit and each is
scaled to the same maximum intensity for illustration purposes
(i.e. the VPs are not normalised). Where it was necessary to bin the
spectra to extract kinematics, the VPs in Fig. \ref{fig:p-v} are
stretched accordingly.  It appears from this figure that the
changes in the stellar velocity and dispersion are synchronised with
the sudden increase in the extinction via the dust lane. However,
Fig. \ref{fig:p-v} suggests that the VPs extracted at the edge of the
dust lane are principally affected at the high-velocity wing; the
low-velocity side of the VP appears unaffected by the presence of the
dust. In fact, if the kinematic fits are extended to include $h_3$ and $h_4$, we find very asymmetric VPs at the location of the dispersion drop with $h_3\sim0.25$.

Further along the slit, we see a very small rise in the dispersion at
the T00 location and the dispersion here is in agreement
with that of the second dispersion peak in T00 of
$\sim70$\kms: there is no clear dispersion peak along this slit
position.

Towards the south end of slit A (at around 340pc on
Fig. \ref{fig:A_avkin}) we see a further rise in \Vs\ and \Ss\
accompanied by another strong gradient in \Vg. The
extinction shows a marginal increase in roughly the same vicinity
(320pc along the slit) but thereafter shows no obvious increase,
although our errors for the extinction are unbound because of the very
small \Ha\ and \Pa\ fluxes received here.


\begin{figure}
    \centering
    \begin{minipage}[c]{0.5\textwidth}
      \includegraphics[width=\textwidth]{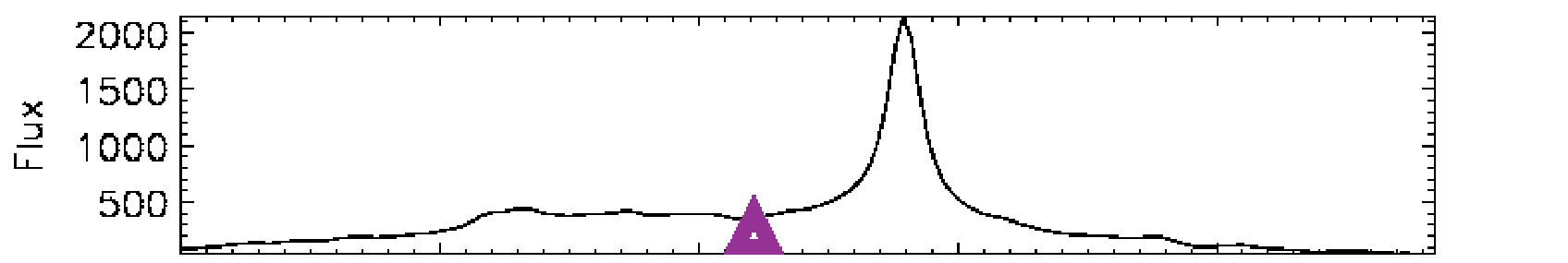}
    \end{minipage}
    \begin{minipage}[c]{0.5\textwidth} 
      \includegraphics[width=\textwidth]{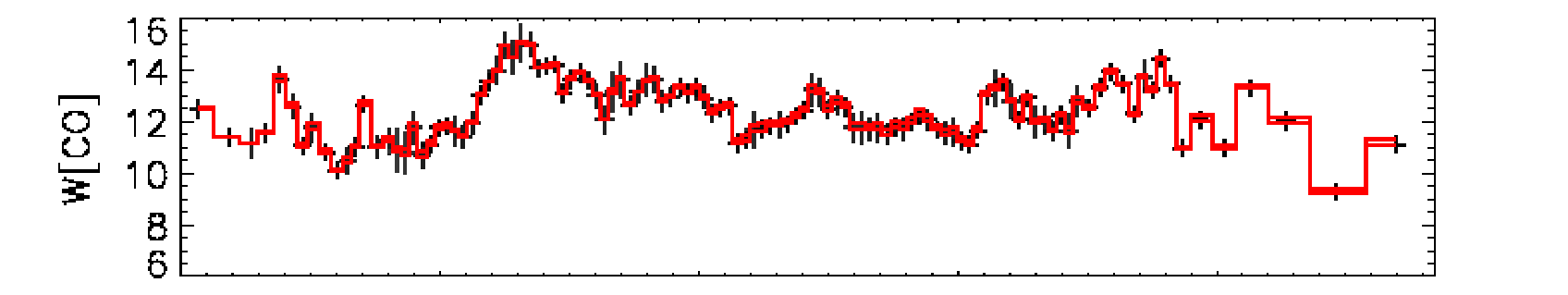}
    \end{minipage}
    \begin{minipage}[c]{0.5\textwidth} 
      \includegraphics[width=\textwidth]{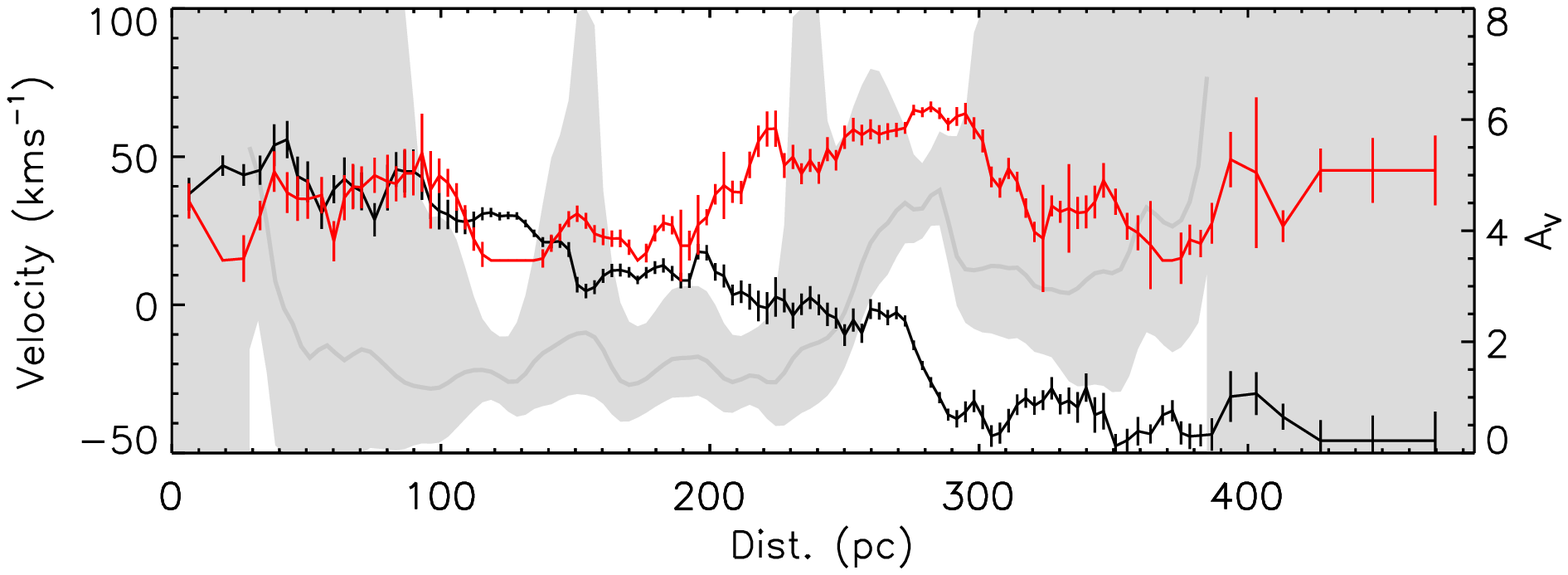}
    \end{minipage}

    \caption{As for Fig. \ref{fig:A_avkin} but for slit E - the
    original data of T00. The position of the putative
    hidden mass concentration from T00 is shown as a
    purple triangle. Note that no H$_2$ gas velocity is shown in the
    lower panel due to its sparse detection along this slit.}
    \label{fig:E_avkin}  
\end{figure}


\begin{table*}
	\vbox to220mm{\vfill  
	Landscape table to go here - 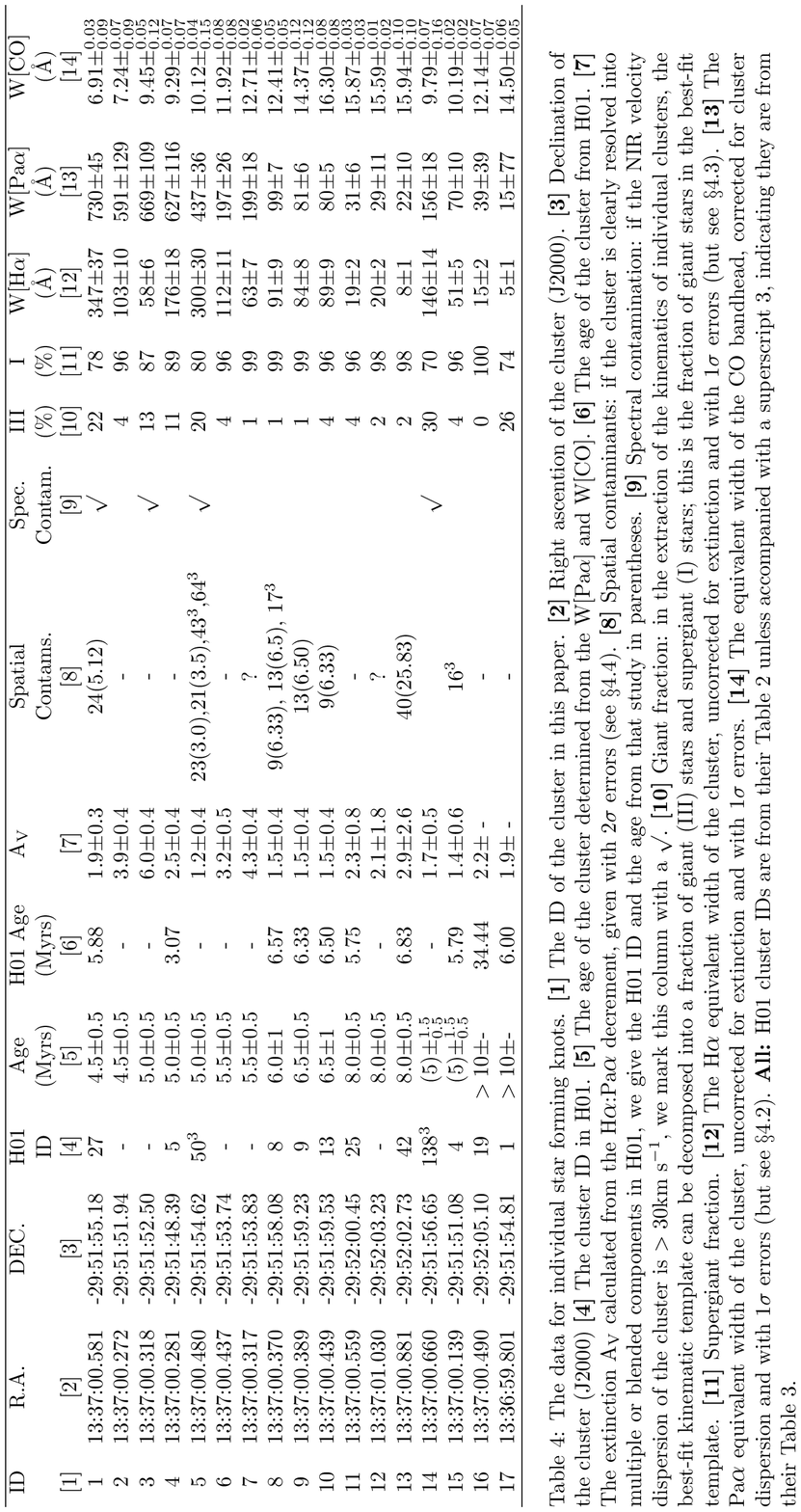
	\vfill}
	\caption{BLANK}
	\label{tab:cluster_ages}
\end{table*}

\section{Discussion}
\label{sec:discussion}

Fig. \ref{fig:cluster_ages} gives the clearest evidence
yet of an age gradient along the circumnuclear arc, with the youngest
clusters nearest (and coincident with) the north-east dust lane. The
ages are generally in agreement with those found by H01, but with
considerably less scatter. As done in D06b, if the H01 data is averaged over angular sections, then the age gradient is very similar to the one found here, although this work extends to the highly extincted clusters in the dust lane and the absolute ages are around a Myr older here.
 
We also see from the false-colour image of Fig. \ref{fig:cluster_ages} that the star
formation does not stop where the north-east dust lane intersects the arc, but
continues through it and along the northern edge: there
appear to be many star clusters at the young end of the arc, embedded
in the dust lane. H01 considered the age gradient along the arc to be
an indication of star formation propagating from the south-east end of
the arc to the north-west, terminating roughly at the dust lane. D06
alternatively proposed that this propagation of star formation was a
consequence of the dynamical friction trailing an interloping hidden
mass concentration, such as a satellite galaxy (although see
\S\ref{sec:dis:noHM}). However, we propose that the arc of young star 
clusters is a result of gas inflow along the north-east dust lane creating a nuclear spiral of gas and an associated ring of star formation, because we find no evidence to support the presence of an interloper at the position of D06(a,b) or the position of T00. 

\subsection{No hidden mass}\label{sec:dis:noHM}

Previous claims of hidden mass concentrations are
inconsistent with our observations for many reasons which we outline below, in addition to the concerns raised in \S\ref{sec:intro:interlopers}. We note that the strong gradient in the molecular gas velocity is 20pc north-west (left along slit A) of the formal position quoted in D06b for the hidden mass concentration. Whether this is a systematic error between coordinate systems or a genuine offset is unclear, although, as discussed in \S\ref{sec:errors:astrometry}, close inspection of the figures in D06b provides a slightly different location, 0\farcs4 (8pc) west of the formal position, going some way to rectifying the difference.

A large mass concentration of the magnitude discussed in D06(a,b) would produce an observable rise in the velocity dispersion --- as seen for the nucleus of similar mass (peak value $\sigma\sim$75\kms). We see no rise in the velocity dispersion at the location of D06 or at the location of the gradient in the molecular gas velocity. The highest stellar velocity dispersion seen at the position of the gradient in molecular gas velocity is $33\pm2$\kms and just 15pc right along slit A is a \emph{minimum} in the stellar velocity dispersion, measured to be $15\pm3$\kms. This minimum in stellar velocity dispersion is probably caused by the young (6 Myrs) star cluster 5 in Table \ref{tab:cluster_ages} but as we shall see, it is highly unlikely that this cluster is dominating the light from an extincted mass concentration. The minimum in stellar velocity dispersion may also be an effect of dust extinction: \citet{Baes03} have shown that dust obscuration can lead to asymmetric velocity profiles and apparent drops in stellar velocity dispersion, although why we see the drop at the location of the shock and not at the peak of the extinction (some 50pc apart) is far from clear in this scenario.

At the location of the putative hidden mass concentration, the extinction (\Av) is 4$\pm$1 (2 sigma errors). This estimate is derived from the HII emission --- the same tracer used to obtain the rotation gradient in D06(a,b). In the K-band, we therefore expect the extinction \Ak\ to be 0.4 magnitudes, so we will receive around two thirds of the total flux emitted from any stars co-spatial with the ionised emission (assuming foreground extinction). Yet we do not see any increase in surface brightness to indicate an increase in stellar density, other than cluster 5, which we measure to have a low stellar velocity dispersion of $\sim15$\kms\ and an age of 6~Myrs. Thus we conclude that there is no large ($\gtsim10^7$\msun) hidden mass there.

We, like H01, observe the youngest star clusters to preceded the trajectory of the putative mass concentration. This is contrary to Ostriker's theory of dynamical friction in a gaseous medium and the subsequent hydrodynamical simulations as discussed in \S\ref{sec:intro:interlopers}. These young clusters, as an extension of the arc, cannot be explained by invoking an interloping mass.

Shocks are well known to give rise to radio emission, as discussed in \S\ref{sec:intro:nuclear_rings} and the MIR is well known to be correlated with the radio; thus the peaks observed in both wavebands near the position of the putative interloper are easily explained by a shock.

As for the other proposed hidden mass concentration, we do not see any prominent change in the stellar velocity dispersion or the molecular gas velocity along slit A or B (both roughly perpendicular to the original data in slit E) at the position proposed by T00: the dispersion peak in the data of T00, shown as slit E here, is therefore most likely a result of combined patchy extinction and young stars leading to a variable stellar velocity dispersion depending on whether the bulge stars or young cluster stars dominate the light.


\subsection{A Nuclear Ring in M83}

Because there is no evidence to support a hidden mass concentration at the centre of M83, we attribute the velocity gradients in the gas (ionised and molecular) to the presence of a shock; in fact, a shock and jump in the gas velocity on the bar's minor axis near the location of the north-east dust lane is predicted by the nuclear ring and spiral simulations of \citet{Regan&Teuben03} and \citet{Witold_paper2}. Furthermore, the simulations of \citeauthor{Witold_paper2} predict shocks on the minor axis that are offset radially inwards to the maximum gas density, which we see is the case in Fig. \ref{fig:A_avkin} if extinction represents gas density. The second jump in the molecular gas velocity, 340pc along slit A, is also likely to be a shock: we can see in Fig. \ref{fig:cluster_ages} that there is a significant quantity of dust 50pc to the south and south-west of the the nucleus, which may be spiralling inwards as in the nuclear spirals of \citet{Witold_paper2}. However, we estimate the position of the kinematic centre defined in \citet{Sakamoto04} to be roughly 240pc along slit A which is almost equidistant from the two gradients in \Vg; therefore, given the scenario proposed by \citet{Sofue&Wakamatsu93} and the symmetry of the CO dust lanes and ring in \citet{Sakamoto04}, it is also possible for this gradient in \Vg\ to emanate from a shock on the inner edge of the south-western dust lane behind the bulge, rather than the aforementioned dust feature. We note that the H$_2$ 2-1 S(1) emission line can be excited both by radiation (UV photons) and collisions (shocks) but it is most intense when collisionally excited (compared to other lines).

The CO bandhead, apart from being able to date the star clusters, also appears to be a useful diagnostic for estimating the \emph{timescale} of the star formation for young clusters ($< 10$Myrs). This
is due to the sharp onset of the feature once the first red
supergiants are formed: \wha\ and \wpa\ evolve relatively slowly in comparison and are not suitable for this purpose (Fig. \ref{fig:sb99eqws} shows that the SSP models and the mixed model predictions for \wha\ and \wpa\ are very similar). As discussed in \S\ref{sec:sb99:mixedmodels}, we measure \wco\ to be consistently larger then the SSP model predictions based on the \wpa\ age. We found that the best way to rectify the conflict between \wco\ and \wpa\ was to replace the instantaneous burst with a finite episode of star formation lasting 6~Myrs. This timescale is consistent with the collapse (and star formation) timescale for a giant molecular cloud with fiducial parameters in the simulations of \citet{Krumholz06}, but it is also comparable to the dynamical timescale for a half revolution around the galaxy centre at this radius. We also tested other finite episodes of star formation with varying star formation rates (exponential growth and exponential decay) and found our data couldn't reliably distinguish between them and the finite episode of constant formation.
The real test of these different star formation scenarios is with clusters $<3$Myrs old, where the model predictions differ most. We also tested a mixed population model of two SSP populations and found that two bursts separated by 5 Myrs with a mass ratio of 1:5 \emph{almost} reproduced the observed data trends in \wpa\ and \wco, but slightly over predicted \wco\ for the youngest clusters. The observation of many more clusters in the north-east dust lane as an extension of the visible arc suggests that the stars are forming inside this dust lane and the mixing timescale of 6~Myrs reflects the formation timescale. However, it is also feasible for the mixing to be a result of multiple age populations. Although we were unable to create multiple burst models that fitted the data as well as the single burst lasting 6~Myrs, the scenario with star formation being triggered in the ring where the dust lanes join (on the minor axis of the bar) is appealing given that it has recently been found in other nuclear rings \citep{Allard06,Sarzi07,FB07}.

\begin{figure}
  \centering
  \includegraphics[width=0.45\textwidth]{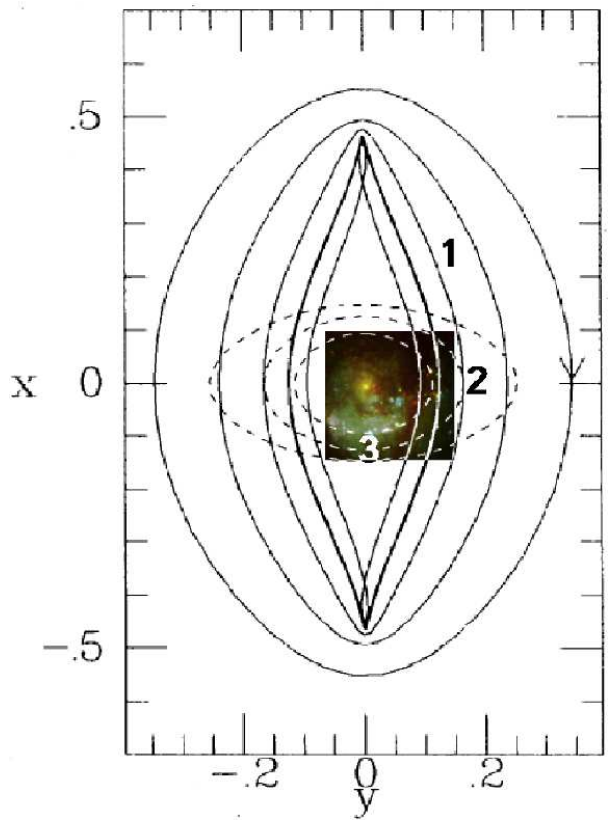}
  \caption{A fiducial illustration of $x_1$ orbits (solid lines) and
  $x_2$ orbits (broken lines) in a typical bar potential (actually for the Milky Way), adapted
  from \citet{Binney91}, with the false-colour image of Fig. 
  \ref{fig:cluster_ages} rotated and superimposed;
  the x-axis is aligned with the bar major axis
  while the y-axis is aligned with the bar minor axis. The bold line shows
  a `cusped' $x_1$ orbit which separates the outer $x_1$
  orbits from their smaller self-intersecting counterparts. Gas
  shocks along the leading edges of the bar and flows inwards to make the dust lane
  (at the point marked 1). This flow gradually funnels the gas onto
  suitable $x_2$ orbits at the point marked 2, although see \citet{Regan&Teuben03} and \citet{Witold_paper2} for complete hydrodynamical simulations. If there was a
  source of star formation at this point (as proposed in the models of \citeauthor{Regan&Teuben03}), such as the partner dust
  lane merging at the same location, then the age gradient along the
  circumnuclear arc could just be a result of the young stars moving
  on $x_2$ orbits away from this location.}
    \label{fig:binney}
\end{figure}

The CO maps of \citet{Sakamoto04} clearly show the molecular hydrogen in both dust
lanes joining in a ring like structure. The material from the
southern dust lane appears to join the ring and move on the opposite side to the
star burst arc to merge with the northern dustlane at
the point where we see the youngest stars. Star formation at this point in the
dust lane would usually be opposed because of the shear forces in action along the
inevitable shocks; however, as in the simulations of \citet{Athanassoula92} and
\citet{Witold_paper2}, we observe the shock front (located by the gradient in
\Vg) on the inner (southern) edge of the dust lane. On the outer edge, away from
the shock front, where there is likely to be spurs and feathering (e.g. as found
in \citealt{BD06aph}) and less shear forces, star formation may be less opposed.
For stars formed here, on the western edge of the dust lane, orbital motion dictates
their trajectory and they will propagate along $x_2$ like orbits (illustrated in Fig. \ref{fig:binney} and also Fig. 15 of \citealt{Regan&Teuben03}) to produce
the `star forming' arc. The gradient in age therefore reflects the older
clusters having propagated further along their mutual $x_2$ orbits.
However, it is not obvious why the youngest star clusters we observe are 4.5~Myrs old; one would expect them to be younger, unless the spraying of material as in \citet{Regan99} led to star formation further up the dust lane. Equally, if star formation occurred on the \emph{opposite side} to where we see the youngest clusters, triggered by the gas in the south-west of the ring merging with the south-west lane (the opposite of the previous scenario), then we would still expect an arc or ring of young stars to form but the timescales (dynamical and population synthesis) would agree better. If this were the case, we presumably do not see the young stars in the north of the ring due to the low luminosity of very young clusters ($<5$~Myrs) and the higher extinction there (seen in Fig. \ref{fig:E_avkin}). Finally, there may be star formation at both sites: the newly formed stars would still propagate on the same $x_2$ orbits but the star formation from each point of contact does not need to be equal: the north-east dustlane appears denser in the CO data of \citeauthor{Sakamoto04} and could dominate the star formation. As previously discussed, we know that dual star formation events, separated by the dynamical timescale, can almost mimic a finite star formation episode of similar duration, but our attempts to fit such a model to the data were not as successfull as a single finite episode. 

A nuclear spiral or ring would cause the young clusters in the arc
to be rotating around the galaxy centre in the same manner:
clockwise as we see them in Fig. \ref{fig:wn_overlay} 
(assuming that the ring is co-planar with the galaxy). M83 is almost
face on, so the rotation is difficult to detect, but the slight
inclination \citep[24\Deg,][]{Talbot79} to our line-of-sight
suggests that in the velocity data of T00, we can see
stars on the south-west side of the nucleus (coincident with the young
clusters) streaming away from us and the stars on the north-east side
stream towards us, in agreement with an inclined clockwise
rotation. However, this slit (labelled slit E in
Fig. \ref{fig:slits_knots}) mainly covers the background population
rather than bright young clusters, so the velocity gradient could
easily represent the overall rotation of the bulge. For consistency,
let us estimate the dynamical time of the stars. From
Fig. \ref{fig:E_avkin}, we see that the average velocity of the stars
either side of the nucleus (and at the position of the arc) is roughly
constant at $\pm25$\kms\ in the rest frame of the galaxy nucleus along
our line of sight. Assuming an inclination of 24\Deg\ then gives us
an estimate for the circular velocity of 60\kms\ $\sim
60$ $\textrm{pc}$ $\textrm{Myr}^{-1}$ (also comparable to the average velocity
dispersion). If the circumnuclear arc is around $15\arcsec \sim
260\textrm{pc}$ in circumference, we expect the stars to take around
4 Myrs to travel from one end to the other. This is in good agreement
with the age decrement found by comparing the cluster ages on either
end of the luminous arc ($\sim$3.5 Myrs). This point was also made in D06b, but the interpretation was significantly different. The dynamical timescale also changes with radius, which may help explain radial age gradients.

We note that the northen dust lane does not appear to follow the same route as the stars along the arc in the HST images but instead continues to propagate south, past the nucleus. This might appear be a concern for the above scenarios. However, the CO maps \citep{Sakamoto04} clearly show the majority of molecular hydrogen gas follows the outer edge of the circumnuclear arc. We also note that in the simulations of \citet{Piner95}, \citet{Regan99} and \citet{Regan&Teuben03}, similar features, where some gas appears to continue past the ring, can sometimes be seen, although they are very faint (not massive): in \citet{Regan99} they are referred to as \emph{spray regions} (also discussed by \citeauthor{Sakamoto04}) and in \citet{Witold_IAUS07} they are seen to be associated with regions of large shear.

\subsection{The displaced nucleus}

If the arc of young stars is a consequence of a nuclear spiral in the gas, we are left pondering why the only visible and kinematically detected nucleus at the centre of M83 is displaced from the centre of the galaxy \emph{and} why we only see an arc of young stars rather than a ring. Apart from the arguments of \citet{Sofue&Wakamatsu93} who suggest that a complete ring is inclined to the galaxy disk and obscured on the northern side, one should also consider a natural m=1 mode. Although rare, M83 is not the only spiral galaxy to show central asymmetries: NGC~1672 is barred spiral with a nucleus displaced from the kinematic and bulge centre, as D06b highlight; NGC~3504 appears to show a single dominant spiral arm in \Ha\ and [NII] encompassing an offset nucleus \citep{Emsellem01b}; NGC~3227 and NGC~7130 are both spirals highlighted in \citet{Munoz07} as having off-centred nuclei; NGC~1808 is another barred spiral with a circumnuclear starburst and nucleus offset from the kinematic centre \citep{TG96,TG05} while NGC~4303 shows asymmetric star formation and dust lanes although the nucleus appears well centred \citep{SMSM02}. 
Although some of these galaxies are clearly interacting, none have clearly merged to an extent where the nuclei are likely to have been accreted externally from the present interaction. \citet{Bournard05} have shown that m=1 modes in the outer regions of disk galaxies can be reproduced with a lopsided external accretion rather than a direct merger: if the outer regions of the galaxy are asymmetric, it may not be easy to define where the true morphological centre lies. That said, M83 does not look particularly disturbed or lopsided in the outer regions. If m=1 modes in the outer disk can be induced by an external accretion event, it may also be possible for m=1 modes to be induced by asymmetric accretion in the inner regions: in the CO maps of \citet{Sakamoto04}, it appears that the gas inflow in M83 is slightly asymmetric, with the north-eastern dust lane denser than the south-west lane at the point of contact to the circumnuclear CO ring; this might cause unequal forcing on any nuclear spiral, although hydrodynamical models are needed to confirm such a possibility. Finally, as noted by \citet{Sakamoto04} the nucleus could be suffering a natural, unprovoked, instability as seen in the models of \citet{TI98} or \citet{Lovelace99} where the central nuclei are able to wander in the disk; such an instability might also be provoked by asymmetric inflow of gas. We noted earlier that the systematic velocity of the gas in the galaxy was taken to be 512 \kms\ but in Fig.~\ref{fig:E_avkin} the  nucleus is moving towards us relative to this frame of reference at around 25\kms, so not only is the nucleus displaced from the centre of the bulge, it is also decoupled kinematically. 

There is no evidence in the stellar kinematics of hidden mass concentrations at the locations previously proposed by T00 or D06(a,b). Without an integral field stellar velocity field of the central region of M83, we cannot completely rule out the presence of another hidden nucleus elsewhere, e.g. near the kinematical and bulge centre. However, given our extinction map (Fig. \ref{fig:ext_map}, from which we estimate \Av\ $< 3$ at the kinematic and bulge centre, corresponding to \Ak$ < 0.3$), the NICMOS F222M photometry (Fig. \ref{fig:wn_overlay}), shows no massive luminous stellar feature other than the well known displaced nucleus: it is difficult to hide $10^7$\msun\ of luminous stars with only 0.3 magnitudes of extinction.

\section{Conclusion}
\label{sec:conc}

We have shown that the NIR indices \wpa\ and \wco\ can be used to
age-date young star clusters. Furthermore, \wpa\ is much less affected
by extinction than \wha, which requires significant correction. The
young clusters in the circumnuclear arc are inconsistent with simple
stellar populations produced by an instantaneous burst; the
observations require a finite episode of star formation lasting around
6 Myrs. Such models together with our observations provide the
clearest evidence yet of an age gradient along the star forming arc,
as first found in H01. However, our NIR observations further show
that the circumnuclear arc does not stop at the edge of the dust lane,
but continues through and along its north-west edge.

Our stellar kinematics show no evidence for a second obscured nucleus
at either the location of D06b or T00 and we
conclude that the velocity gradients in the gas kinematics (ionised and molecular) are a
result of a shock front on the south-eastern edge of the north-eastern dust lane. We
therefore explore alternative explanations for the formation of the
circumnuclear arc, the most plausible being the $x_2$ orbital
motion of the clusters from their place of birth in the dust lane.

Future higher-resolution integral field spectroscopy in the NIR has
the potential to build on this study and find the youngest clusters in
the arc (suspected to be on the north-west edge of the northern dust lane) and
also address previous claims of an age gradient \emph{perpendicular}
to the arc (H01). 

\section{Acknowledgements}
\label{sec:ack}
We wish to thank the referee for his/her constructive comments and Sabine Mengel for kindly providing the supergiant templates which we use when extracting the kinematics. We are also grateful for many fruitfull discussions with Witold Maciejewski and James Binney. We acknowledge that the linear sequence of radio sources proposed to be a background galaxy by \citet{Maddox06} was discussed in a talk by \citet[][proceedings currently unpublished and no pre-print available]{Dottori07} as a feature in the M83 disk possibly linked to the offset nucleus.
This research is based on observations collected at the European Southern
Observatory and the Hubble Space Telescope. We acknowledge use of the
SIMBAD Astronomical Database, the HyperLeda database and NASA's IDL
Astronomy User's Library. RH is funded by the University of Oxford and NT is funded through the Marie-Curie Excellence Grant MEXT-CT-2003-002792.

\bibliographystyle{mn2e}
\bibliography{HOUGHTON_M83}

\begin{thebibliography}{89}
\expandafter\ifx\csname natexlab\endcsname\relax\def\natexlab#1{#1}\fi

\bibitem[{{Allard} {et~al.}(2006){Allard}, {Knapen}, {Peletier}, \&
  {Sarzi}}]{Allard06}
{Allard} E.~L., {Knapen} J.~H., {Peletier} R.~F., {Sarzi} M., 2006, \mnras,
  371, 1087

\bibitem[{{Athanassoula}(1992)}]{Athanassoula92}
{Athanassoula} E., 1992, \mnras, 259, 345

\bibitem[{{Baes} {et~al.}(2003){Baes}, {Davies}, {Dejonghe}, {Sabatini},
  {Roberts}, {Evans}, {Linder}, {Smith}, \& {de Blok}}]{Baes03}
{Baes} M., {Davies} J.~I., {Dejonghe} H., {Sabatini} S., {Roberts} S., {Evans}
  R., {Linder} S.~M., {Smith} R.~M., {de Blok} W.~J.~G., 2003, \mnras, 343,
  1081

\bibitem[{{Binney} {et~al.}(1991){Binney}, {Gerhard}, {Stark}, {Bally}, \&
  {Uchida}}]{Binney91}
{Binney} J., {Gerhard} O.~E., {Stark} A.~A., {Bally} J., {Uchida} K.~I., 1991,
  \mnras, 252, 210

\bibitem[{{Bonnell} \& {Dobbs}(2006)}]{BD06aph}
{Bonnell} I.~A., {Dobbs} C.~L., 2006, ArXiv Astrophysics e-prints

\bibitem[{{Bournaud} {et~al.}(2005){Bournaud}, {Combes}, {Jog}, \&
  {Puerari}}]{Bournard05}
{Bournaud} F., {Combes} F., {Jog} C.~J., {Puerari} I., 2005, \aap, 438, 507

\bibitem[{{Bresolin} \& {Kennicutt}(2002)}]{Bresolin&Kennicutt03}
{Bresolin} F., {Kennicutt} Jr. R.~C., 2002, \apj, 572, 838

\bibitem[{{Buta}(1986)}]{Buta86}
{Buta} R., 1986, \apjs, 61, 609

\bibitem[{{Buta} \& {Combes}(1996)}]{Buta&Combes96}
{Buta} R., {Combes} F., 1996, Fundamentals of Cosmic Physics, 17, 95

\bibitem[{{Calzetti}(1997)}]{Calzetti97}
{Calzetti} D., 1997, \aj, 113, 162

\bibitem[{{Calzetti}(2001)}]{Calzetti01}
---, 2001, \pasp, 113, 1449

\bibitem[{{Calzetti} {et~al.}(2004){Calzetti}, {Harris}, {Gallagher}, {Smith},
  {Conselice}, {Homeier}, \& {Kewley}}]{Calzetti04}
{Calzetti} D., {Harris} J., {Gallagher} III J.~S., {Smith} D.~A., {Conselice}
  C.~J., {Homeier} N., {Kewley} L., 2004, \aj, 127, 1405

\bibitem[{{Calzetti} {et~al.}(1994){Calzetti}, {Kinney}, \&
  {Storchi-Bergmann}}]{Calzetti94}
{Calzetti} D., {Kinney} A.~L., {Storchi-Bergmann} T., 1994, \apj, 429, 582

\bibitem[{{Cardelli} {et~al.}(1989){Cardelli}, {Clayton}, \&
  {Mathis}}]{Cardelli89}
{Cardelli} J.~A., {Clayton} G.~C., {Mathis} J.~S., 1989, \apj, 345, 245

\bibitem[{{Chung} \& {Bureau}(2004)}]{Chung&Bureau04}
{Chung} A., {Bureau} M., 2004, \aj, 127, 3192

\bibitem[{{Combes}(1996)}]{Combes96}
{Combes} F., 1996, in Astronomical Society of the Pacific Conference Series,
  Vol.~91, IAU Colloq. 157: Barred Galaxies, {Buta} R., {Crocker} D.~A.,
  {Elmegreen} B.~G., eds., pp. 286--+

\bibitem[{{Cowan} {et~al.}(1994){Cowan}, {Roberts}, \& {Branch}}]{Cowan94}
{Cowan} J.~J., {Roberts} D.~A., {Branch} D., 1994, \apj, 434, 128

\bibitem[{{de Vaucouleurs} {et~al.}(1991){de Vaucouleurs}, {de Vaucouleurs},
  {Corwin}, {Buta}, {Paturel}, \& {Fouque}}]{deVauc91}
{de Vaucouleurs} G., {de Vaucouleurs} A., {Corwin} Jr. H.~G., {Buta} R.~J.,
  {Paturel} G., {Fouque} P., 1991, {Third Reference Catalogue of Bright
  Galaxies}. Volume 1-3, XII, 2069 pp.~7 figs..~ Springer-Verlag Berlin
  Heidelberg New York

\bibitem[{{Diaz} {et~al.}(2007){Diaz}, {Rodrigues}, {Dottori}, {Mast}, \&
  {Ag{\"u}ero}}]{Diaz07}
{Diaz} R., {Rodrigues} I., {Dottori} H., {Mast} D., {Ag{\"u}ero} M.~P., 2007,
  in IAU Symposium, Vol. 235, IAU Symposium, {Combes} F., {Palous} J., eds.,
  pp. 93--93

\bibitem[{{D{\'{\i}}az} {et~al.}(2006b){D{\'{\i}}az}, {Dottori}, {Aguero},
  {Mediavilla}, {Rodrigues}, \& {Mast}}]{Diaz06b}
{D{\'{\i}}az} R.~J., {Dottori} H., {Aguero} M.~P., {Mediavilla} E., {Rodrigues}
  I., {Mast} D., 2006b, \apj, 652, 1122

\bibitem[{{Diaz} {et~al.}(2006a){Diaz}, {Dottori}, {Mediavilla}, {Aguero}, \&
  {Mast}}]{Diaz06a}
{Diaz} R.~J., {Dottori} H., {Mediavilla} E., {Aguero} M., {Mast} D., 2006a, New
  Astronomy Review, 49, 547

\bibitem[{{Dottori} {et~al.}(2007){Dottori}, {Diaz}, {Aguero}, {Mast}, \&
  {Rodrigues}}]{Dottori07}
{Dottori} H., {Diaz} R.~J., {Aguero} M.~P., {Mast} D., {Rodrigues} I., 2007, in
  IAU Symposium, Vol. 245, IAU Symposium, in press, {Bureau} M., ed.

\bibitem[{{Elmegreen} {et~al.}(1998){Elmegreen}, {Chromey}, \&
  {Warren}}]{Elmegreen98}
{Elmegreen} D.~M., {Chromey} F.~R., {Warren} A.~R., 1998, \aj, 116, 2834

\bibitem[{{Emsellem}(2001)}]{Emsellem01b}
{Emsellem} E., 2001, in Astronomical Society of the Pacific Conference Series,
  Vol. 249, The Central Kiloparsec of Starbursts and AGN: The La Palma
  Connection, {Knapen} J.~H., {Beckman} J.~E., {Shlosman} I., {Mahoney} T.~J.,
  eds., pp. 91--+

\bibitem[{{Emsellem} {et~al.}(2001){Emsellem}, {Greusard}, {Combes}, {Friedli},
  {Leon}, {P{\'e}contal}, \& {Wozniak}}]{Emsellem01}
{Emsellem} E., {Greusard} D., {Combes} F., {Friedli} D., {Leon} S.,
  {P{\'e}contal} E., {Wozniak} H., 2001, \aap, 368, 52

\bibitem[{{Escala} {et~al.}(2004){Escala}, {Larson}, {Coppi}, \&
  {Mardones}}]{Escala04}
{Escala} A., {Larson} R.~B., {Coppi} P.~S., {Mardones} D., 2004, \apj, 607, 765

\bibitem[{{Fagotto} {et~al.}(1994){Fagotto}, {Bressan}, {Bertelli}, \&
  {Chiosi}}]{Padova}
{Fagotto} F., {Bressan} A., {Bertelli} G., {Chiosi} C., 1994, \aaps, 105, 29

\bibitem[{{Falc{\'o}n-Barroso} {et~al.}(2007){Falc{\'o}n-Barroso}, {Boeker},
  {Schinnerer}, {Knapen}, \& {Ryder}}]{FB07}
{Falc{\'o}n-Barroso} J., {Boeker} T., {Schinnerer} E., {Knapen} J.~H., {Ryder}
  S., 2007, ArXiv e-prints, 709

\bibitem[{{F{\"o}rster Schreiber} {et~al.}(2003){F{\"o}rster Schreiber},
  {Genzel}, {Lutz}, \& {Sternberg}}]{FS03}
{F{\"o}rster Schreiber} N.~M., {Genzel} R., {Lutz} D., {Sternberg} A., 2003,
  \apj, 599, 193

\bibitem[{{Gallais} {et~al.}(1991){Gallais}, {Rouan}, {Lacombe}, {Tiphene}, \&
  {Vauglin}}]{Gallais91}
{Gallais} P., {Rouan} D., {Lacombe} F., {Tiphene} D., {Vauglin} I., 1991, \aap,
  243, 309

\bibitem[{{Harris} {et~al.}(2001){Harris}, {Calzetti}, {Gallagher},
  {Conselice}, \& {Smith}}]{Harris01}
{Harris} J., {Calzetti} D., {Gallagher} III J.~S., {Conselice} C.~J., {Smith}
  D.~A., 2001, \aj, 122, 3046

\bibitem[{{Heap} {et~al.}(1993){Heap}, {Holbrook}, {Malumuth}, {Shore}, \&
  {Waller}}]{Heap93}
{Heap} S.~R., {Holbrook} J., {Malumuth} E., {Shore} S., {Waller} W., 1993, in
  Bulletin of the American Astronomical Society, Vol.~25, Bulletin of the
  American Astronomical Society, pp. 840--+

\bibitem[{{Hinkle} {et~al.}(2000){Hinkle}, {Aringer}, {Lebzelter}, {Martin}, \&
  {Ridgway}}]{H22-1S1}
{Hinkle} K.~H., {Aringer} B., {Lebzelter} T., {Martin} C.~L., {Ridgway} S.~T.,
  2000, \aap, 363, 1065

\bibitem[{{Holtzman} {et~al.}(1995){Holtzman}, {Hester}, {Casertano},
  {Trauger}, {Watson}, {Ballester}, {Burrows}, {Clarke}, {Crisp}, {Evans},
  {Gallagher}, {Griffiths}, {Hoessel}, {Matthews}, {Mould}, {Scowen},
  {Stapelfeldt}, \& {Westphal}}]{Holtzman95}
{Holtzman} J.~A., {Hester} J.~J., {Casertano} S., {Trauger} J.~T., {Watson}
  A.~M., {Ballester} G.~E., {Burrows} C.~J., {Clarke} J.~T., {Crisp} D.,
  {Evans} R.~W., {Gallagher} III J.~S., {Griffiths} R.~E., {Hoessel} J.~G.,
  {Matthews} L.~D., {Mould} J.~R., {Scowen} P.~A., {Stapelfeldt} K.~R.,
  {Westphal} J.~A., 1995, \pasp, 107, 156

\bibitem[{{Houghton} {et~al.}(2006){Houghton}, {Magorrian}, {Sarzi}, {Thatte},
  {Davies}, \& {Krajnovi{\'c}}}]{Houghton06}
{Houghton} R.~C.~W., {Magorrian} J., {Sarzi} M., {Thatte} N., {Davies} R.~L.,
  {Krajnovi{\'c}} D., 2006, \mnras, 367, 2

\bibitem[{{Kenney} {et~al.}(1993){Kenney}, {Carlstrom}, \&
  {Young}}]{KenneyCarlstromYoung93}
{Kenney} J.~D.~P., {Carlstrom} J.~E., {Young} J.~S., 1993, \apj, 418, 687

\bibitem[{{Kim} \& {Kim}(2007)}]{Kim&Kim07}
{Kim} H., {Kim} W.-T., 2007, \apj, 665, 432

\bibitem[{{Kleinmann} \& {Hall}(1986)}]{KH86}
{Kleinmann} S.~G., {Hall} D.~N.~B., 1986, \apjs, 62, 501

\bibitem[{{Kobulnicky} {et~al.}(1999){Kobulnicky}, {Kennicutt}, \&
  {Pizagno}}]{METAL}
{Kobulnicky} H.~A., {Kennicutt} Jr. R.~C., {Pizagno} J.~L., 1999, \apj, 514,
  544

\bibitem[{{Koribalski} {et~al.}(2004){Koribalski}, {Staveley-Smith}, {Kilborn},
  {Ryder}, {Kraan-Korteweg}, {Ryan-Weber}, {Ekers}, {Jerjen}, {Henning},
  {Putman}, {Zwaan}, {de Blok}, {Calabretta}, {Disney}, {Minchin}, {Bhathal},
  {Boyce}, {Drinkwater}, {Freeman}, {Gibson}, {Green}, {Haynes}, {Juraszek},
  {Kesteven}, {Knezek}, {Mader}, {Marquarding}, {Meyer}, {Mould}, {Oosterloo},
  {O'Brien}, {Price}, {Sadler}, {Schr{\"o}der}, {Stewart}, {Stootman}, {Waugh},
  {Warren}, {Webster}, \& {Wright}}]{Kori04}
{Koribalski} B.~S., {Staveley-Smith} L., {Kilborn} V.~A., {Ryder} S.~D.,
  {Kraan-Korteweg} R.~C., {Ryan-Weber} E.~V., {Ekers} R.~D., {Jerjen} H.,
  {Henning} P.~A., {Putman} M.~E., {Zwaan} M.~A., {de Blok} W.~J.~G.,
  {Calabretta} M.~R., {Disney} M.~J., {Minchin} R.~F., {Bhathal} R., {Boyce}
  P.~J., {Drinkwater} M.~J., {Freeman} K.~C., {Gibson} B.~K., {Green} A.~J.,
  {Haynes} R.~F., {Juraszek} S., {Kesteven} M.~J., {Knezek} P.~M., {Mader} S.,
  {Marquarding} M., {Meyer} M., {Mould} J.~R., {Oosterloo} T., {O'Brien} J.,
  {Price} R.~M., {Sadler} E.~M., {Schr{\"o}der} A., {Stewart} I.~M., {Stootman}
  F., {Waugh} M., {Warren} B.~E., {Webster} R.~L., {Wright} A.~E., 2004, \aj,
  128, 16

\bibitem[{{Kormendy} \& {Richstone}(1995)}]{KormRich95}
{Kormendy} J., {Richstone} D., 1995, \araa, 33, 581

\bibitem[{{Krumholz} {et~al.}(2006){Krumholz}, {Matzner}, \&
  {McKee}}]{Krumholz06}
{Krumholz} M.~R., {Matzner} C.~D., {McKee} C.~F., 2006, \apj, 653, 361

\bibitem[{{Larsen} \& {Richtler}(1999)}]{Larsen99}
{Larsen} S.~S., {Richtler} T., 1999, \aap, 345, 59

\bibitem[{{Leitherer} \& {Heckman}(1995)}]{LH95}
{Leitherer} C., {Heckman} T.~M., 1995, \apjs, 96, 9

\bibitem[{{Leitherer} {et~al.}(1999){Leitherer}, {Schaerer}, {Goldader},
  {Delgado}, {Robert}, {Kune}, {de Mello}, {Devost}, \& {Heckman}}]{SB99}
{Leitherer} C., {Schaerer} D., {Goldader} J.~D., {Delgado} R.~M.~G., {Robert}
  C., {Kune} D.~F., {de Mello} D.~F., {Devost} D., {Heckman} T.~M., 1999,
  \apjs, 123, 3

\bibitem[{{Lovelace} {et~al.}(1999){Lovelace}, {Zhang}, {Kornreich}, \&
  {Haynes}}]{Lovelace99}
{Lovelace} R.~V.~E., {Zhang} L., {Kornreich} D.~A., {Haynes} M.~P., 1999, \apj,
  524, 634

\bibitem[{{Maciejewski}(2004)}]{Witold_paper2}
{Maciejewski} W., 2004, \mnras, 354, 892

\bibitem[{{Maciejewski}(2007)}]{Witold_IAUS07}
---, 2007, ArXiv e-prints, 709

\bibitem[{{Maddox} {et~al.}(2006){Maddox}, {Cowan}, {Kilgard}, {Lacey},
  {Prestwich}, {Stockdale}, \& {Wolfing}}]{Maddox06}
{Maddox} L.~A., {Cowan} J.~J., {Kilgard} R.~E., {Lacey} C.~K., {Prestwich}
  A.~H., {Stockdale} C.~J., {Wolfing} E., 2006, \aj, 132, 310

\bibitem[{{Malin} \& {Hadley}(1997)}]{Malin97}
{Malin} D., {Hadley} B., 1997, Publications of the Astronomical Society of
  Australia, 14, 52

\bibitem[{{Martini} \& {Pogge}(1999)}]{MP99}
{Martini} P., {Pogge} R.~W., 1999, \aj, 118, 2646

\bibitem[{{Mast} {et~al.}(2006){Mast}, {D{\'{\i}}az}, \& {Ag{\"u}ero}}]{Mast06}
{Mast} D., {D{\'{\i}}az} R.~J., {Ag{\"u}ero} M.~P., 2006, \aj, 131, 1394

\bibitem[{{Meynet} {et~al.}(1994){Meynet}, {Maeder}, {Schaller}, {Schaerer}, \&
  {Charbonnel}}]{Geneva}
{Meynet} G., {Maeder} A., {Schaller} G., {Schaerer} D., {Charbonnel} C., 1994,
  \aaps, 103, 97

\bibitem[{{Miller} \& {Bregman}(2005)}]{Miller&Bregman04}
{Miller} E.~D., {Bregman} J.~N., 2005, in Astronomical Society of the Pacific
  Conference Series, Vol. 331, Extra-Planar Gas, {Braun} R., ed., pp. 261--+

\bibitem[{{Mu{\~n}oz Mar{\'{\i}}n} {et~al.}(2007){Mu{\~n}oz Mar{\'{\i}}n},
  {Gonz{\'a}lez Delgado}, {Schmitt}, {Cid Fernandes}, {P{\'e}rez},
  {Storchi-Bergmann}, {Heckman}, \& {Leitherer}}]{Munoz07}
{Mu{\~n}oz Mar{\'{\i}}n} V.~M., {Gonz{\'a}lez Delgado} R.~M., {Schmitt} H.~R.,
  {Cid Fernandes} R., {P{\'e}rez} E., {Storchi-Bergmann} T., {Heckman} T.,
  {Leitherer} C., 2007, \aj, 134, 648

\bibitem[{{Oliva} {et~al.}(1995){Oliva}, {Origlia}, {Kotilainen}, \&
  {Moorwood}}]{ookm95}
{Oliva} E., {Origlia} L., {Kotilainen} J.~K., {Moorwood} A.~F.~M., 1995, \aap,
  301, 55

\bibitem[{{Origlia} {et~al.}(1999){Origlia}, {Goldader}, {Leitherer},
  {Schaerer}, \& {Oliva}}]{Origlia99}
{Origlia} L., {Goldader} J.~D., {Leitherer} C., {Schaerer} D., {Oliva} E.,
  1999, \apj, 514, 96

\bibitem[{{Origlia} {et~al.}(1993){Origlia}, {Moorwood}, \& {Oliva}}]{omo93}
{Origlia} L., {Moorwood} A.~F.~M., {Oliva} E., 1993, \aap, 280, 536

\bibitem[{{Osterbrock}(1989)}]{Osterbrock89}
{Osterbrock} D.~E., 1989, {Astrophysics of gaseous nebulae and active galactic
  nuclei}. Research supported by the University of California, John Simon
  Guggenheim Memorial Foundation, University of Minnesota, et al.~Mill Valley,
  CA, University Science Books, 1989, 422 p.

\bibitem[{{Ostriker}(1999)}]{Ostriker99}
{Ostriker} E.~C., 1999, \apj, 513, 252

\bibitem[{{Park} {et~al.}(2001){Park}, {Kalnajs}, {Freeman}, {Koribalski},
  {Staveley-Smith}, \& {Malin}}]{Park01}
{Park} O.-K., {Kalnajs} A., {Freeman} K.~C., {Koribalski} B., {Staveley-Smith}
  L., {Malin} D.~F., 2001, in Astronomical Society of the Pacific Conference
  Series, Vol. 230, Galaxy Disks and Disk Galaxies, {Funes} J.~G., {Corsini}
  E.~M., eds., pp. 109--110

\bibitem[{{Piner} {et~al.}(1995){Piner}, {Stone}, \& {Teuben}}]{Piner95}
{Piner} B.~G., {Stone} J.~M., {Teuben} P.~J., 1995, \apj, 449, 508

\bibitem[{{Pogge} \& {Martini}(2002)}]{PM02}
{Pogge} R.~W., {Martini} P., 2002, \apj, 569, 624

\bibitem[{{Puxley} {et~al.}(1997){Puxley}, {Doyon}, \& {Ward}}]{Pux97}
{Puxley} P.~J., {Doyon} R., {Ward} M.~J., 1997, \apj, 476, 120

\bibitem[{{Regan} {et~al.}(1999){Regan}, {Sheth}, \& {Vogel}}]{Regan99}
{Regan} M.~W., {Sheth} K., {Vogel} S.~N., 1999, \apj, 526, 97

\bibitem[{{Regan} \& {Teuben}(2003)}]{Regan&Teuben03}
{Regan} M.~W., {Teuben} P., 2003, \apj, 582, 723

\bibitem[{{Rix} \& {White}(1992)}]{RW92}
{Rix} H.-W., {White} S.~D.~M., 1992, \mnras, 254, 389

\bibitem[{{Rogstad} {et~al.}(1974){Rogstad}, {Lockart}, \& {Wright}}]{Rog74}
{Rogstad} D.~H., {Lockart} I.~A., {Wright} M.~C.~H., 1974, \apj, 193, 309

\bibitem[{{Ryder} {et~al.}(2005){Ryder}, {Sharp}, {Knapen}, {Mazzuca}, \&
  {Parry}}]{Ryder05}
{Ryder} S.~D., {Sharp} R.~G., {Knapen} J.~H., {Mazzuca} L.~M., {Parry} I.~R.,
  2005, in American Institute of Physics Conference Series, Vol. 783, The
  Evolution of Starbursts, {H{\"u}ttmeister} S., {Manthey} E., {Bomans} D.,
  {Weis} K., eds., pp. 155--160

\bibitem[{{Sakamoto} {et~al.}(2004){Sakamoto}, {Matsushita}, {Peck}, {Wiedner},
  \& {Iono}}]{Sakamoto04}
{Sakamoto} K., {Matsushita} S., {Peck} A.~B., {Wiedner} M.~C., {Iono} D., 2004,
  \apjl, 616, L59

\bibitem[{{Sandage} \& {Bedke}(1994)}]{CarnegieAtlas}
{Sandage} A., {Bedke} J., 1994, {The Carnegie atlas of galaxies}. Washington,
  DC: Carnegie Institution of Washington with The Flintridge Foundation, |c1994

\bibitem[{{Sarazin} \& {Roddier}(1990)}]{ESODIMM}
{Sarazin} M., {Roddier} F., 1990, \aap, 227, 294

\bibitem[{{Sarzi} {et~al.}(2007){Sarzi}, {Allard}, {Knapen}, \&
  {Mazzuca}}]{Sarzi07}
{Sarzi} M., {Allard} E.~L., {Knapen} J.~H., {Mazzuca} L.~M., 2007, \mnras, 380,
  949

\bibitem[{{Sarzi} {et~al.}(2006){Sarzi}, {Falc{\'o}n-Barroso}, {Davies},
  {Bacon}, {Bureau}, {Cappellari}, {de Zeeuw}, {Emsellem}, {Fathi},
  {Krajnovi{\'c}}, {Kuntschner}, {McDermid}, \& {Peletier}}]{Sarzi06}
{Sarzi} M., {Falc{\'o}n-Barroso} J., {Davies} R.~L., {Bacon} R., {Bureau} M.,
  {Cappellari} M., {de Zeeuw} P.~T., {Emsellem} E., {Fathi} K., {Krajnovi{\'c}}
  D., {Kuntschner} H., {McDermid} R.~M., {Peletier} R.~F., 2006, \mnras, 366,
  1151

\bibitem[{{Schinnerer} {et~al.}(2002){Schinnerer}, {Maciejewski}, {Scoville},
  \& {Moustakas}}]{SMSM02}
{Schinnerer} E., {Maciejewski} W., {Scoville} N., {Moustakas} L.~A., 2002,
  \apj, 575, 826

\bibitem[{{Sofue} \& {Wakamatsu}(1994)}]{Sofue&Wakamatsu93}
{Sofue} Y., {Wakamatsu} K.-I., 1994, \aj, 107, 1018

\bibitem[{{Soria} \& {Wu}(2003)}]{Soria&Wu03}
{Soria} R., {Wu} K., 2003, \aap, 410, 53

\bibitem[{{Storchi-Bergmann} {et~al.}(1995){Storchi-Bergmann}, {Kinney}, \&
  {Challis}}]{Storchi95}
{Storchi-Bergmann} T., {Kinney} A.~L., {Challis} P., 1995, \apjs, 98, 103

\bibitem[{{Tacconi-Garman} {et~al.}(1996){Tacconi-Garman}, {Sternberg}, \&
  {Eckart}}]{TG96}
{Tacconi-Garman} L.~E., {Sternberg} A., {Eckart} A., 1996, \aj, 112, 918

\bibitem[{{Tacconi-Garman} {et~al.}(2005){Tacconi-Garman}, {Sturm}, {Lehnert},
  {Lutz}, {Davies}, \& {Moorwood}}]{TG05}
{Tacconi-Garman} L.~E., {Sturm} E., {Lehnert} M., {Lutz} D., {Davies} R.~I.,
  {Moorwood} A.~F.~M., 2005, \aap, 432, 91

\bibitem[{{Taga} \& {Iye}(1998)}]{TI98}
{Taga} M., {Iye} M., 1998, \mnras, 299, 1132

\bibitem[{{Talbot} {et~al.}(1979){Talbot}, {Jensen}, \& {Dufour}}]{Talbot79}
{Talbot} Jr. R.~J., {Jensen} E.~B., {Dufour} R.~J., 1979, \apj, 229, 91

\bibitem[{{Thatte} {et~al.}(2000){Thatte}, {Tecza}, \& {Genzel}}]{Thatte00}
{Thatte} N., {Tecza} M., {Genzel} R., 2000, \aap, 364, L47

\bibitem[{{Thilker} {et~al.}(2005){Thilker}, {Bianchi}, {Boissier}, {Gil de
  Paz}, {Madore}, {Martin}, {Meurer}, {Neff}, {Rich}, {Schiminovich},
  {Seibert}, {Wyder}, {Barlow}, {Byun}, {Donas}, {Forster}, {Friedman},
  {Heckman}, {Jelinsky}, {Lee}, {Malina}, {Milliard}, {Morrissey}, {Siegmund},
  {Small}, {Szalay}, \& {Welsh}}]{Thilker05}
{Thilker} D.~A., {Bianchi} L., {Boissier} S., {Gil de Paz} A., {Madore} B.~F.,
  {Martin} D.~C., {Meurer} G.~R., {Neff} S.~G., {Rich} R.~M., {Schiminovich}
  D., {Seibert} M., {Wyder} T.~K., {Barlow} T.~A., {Byun} Y.-I., {Donas} J.,
  {Forster} K., {Friedman} P.~G., {Heckman} T.~M., {Jelinsky} P.~N., {Lee}
  Y.-W., {Malina} R.~F., {Milliard} B., {Morrissey} P., {Siegmund} O.~H.~W.,
  {Small} T., {Szalay} A.~S., {Welsh} B.~Y., 2005, \apjl, 619, L79

\bibitem[{{Thim} {et~al.}(2003){Thim}, {Tammann}, {Saha}, {Dolphin}, {Sandage},
  {Tolstoy}, \& {Labhardt}}]{Thim03}
{Thim} F., {Tammann} G.~A., {Saha} A., {Dolphin} A., {Sandage} A., {Tolstoy}
  E., {Labhardt} L., 2003, \apj, 590, 256

\bibitem[{{V{\'a}zquez} \& {Leitherer}(2005)}]{SB99update}
{V{\'a}zquez} G.~A., {Leitherer} C., 2005, \apj, 621, 695

\bibitem[{{Wolstencroft}(1988)}]{Wolst88}
{Wolstencroft} R.~., 1988, in Proceedings of a Conference, held in honor of the
  60th birthday of Halton C. Arp, Venice, Italy, May 5-7, 1987, Cambridge:
  University Press, 1988, edited by Bertola, F.; Sulentic, J.W.; Madore, B.F.,
  {Bertola} F., {Sulentic} J., {Madore} B., eds.

\bibitem[{{Wozniak} \& {Champavert}(2006)}]{Wozniak&Champavert06}
{Wozniak} H., {Champavert} N., 2006, \mnras, 369, 853

\bibitem[{{Wozniak} {et~al.}(2003){Wozniak}, {Combes}, {Emsellem}, \&
  {Friedli}}]{Wozniak03}
{Wozniak} H., {Combes} F., {Emsellem} E., {Friedli} D., 2003, \aap, 409, 469

\end{thebibliography}

\label{lastpage}
\end{document}